\documentclass{elsart}

\usepackage{natbib}

\usepackage{graphics}
\usepackage{graphicx}
 \usepackage{epsfig}

\usepackage{amssymb}

\newcommand{\beq}{\begin{equation}}
\newcommand{\eeq}{\end{equation}}

\newcommand{\D}{\Delta}
\newcommand{\p}{\partial}

\newcommand{\bea}{\begin{eqnarray}}
\newcommand{\eea}{\end{eqnarray}}

\begin{document}

\begin{frontmatter}


 \thanks[label3]{In the memory of our friend and colleague Simos Ichtiaroglou}

\title{Normal and Anomalous Diffusion: A Tutorial}

\author{Loukas Vlahos, Heinz Isliker}
\address{Department of Physics, University of Thessaloniki,
54124 Thessaloniki, Greece}
\author{Yannis Kominis, and Kyriakos Hizanidis}
\address{School of Electrical and Computer Engineering, National Technical University of Athens,
15773 Zografou, Athens, Greece}

\begin{abstract}
The purpose of this tutorial is to introduce the main concepts behind
normal and anomalous diffusion. Starting from simple, but well known
experiments, a series of mathematical modeling tools are introduced,
and the relation between them is made clear. First, we show how Brownian motion can be understood
in terms of a simple random walk model. Normal diffusion is then
treated (i) through formalizing the random walk model and deriving a
classical diffusion equation, (ii) by using Fick's law that leads
again to the same diffusion equation, and (iii) by using a stochastic
differential equation for the particle dynamics (the Langevin equation),
which allows to determine the mean square displacement of particles.
(iv) We discuss normal diffusion from the point of view of probability
theory, applying the Central Limit Theorem to the random walk problem,
and (v) we introduce the more general Fokker-Planck equation
for diffusion that includes also advection. We turn then to anomalous
diffusion, discussing first its formal characteristics, and proceeding
to Continuous Time Random Walk (CTRW) as a model for anomalous diffusion.
It is shown how CTRW can be treated formally, the importance of probability
distributions of the Levy type is explained,
and we discuss the relation of CTRW to fractional diffusion equations and
show how the latter can be derived from the CTRW equations.
Last, we demonstrate how a general diffusion equation can be derived
for Hamiltonian systems, and
we conclude this tutorial with a few recent
applications of the above theories in laboratory and astrophysical plasmas.
\end{abstract}


\end{frontmatter}


\section{Introduction}

\label{intro}
The art of doing research in physics  usually starts with the  observation of
a natural
phenomenon. Then follows a qualitative idea on "How the phenomenon can be
interpreted",
and one proceeds with the construction of a model equation or a simulation,
with the aim that it resembles
very well the observed phenomenon.  This progression from natural phenomena
to models
and mathematical prototypes and then back to many similar natural phenomena,
is the methodological beauty of our research in physics.

Diffusion belongs to this class of phenomena. All started from the
observations of several
scientists on the irregular motion of dust, coal or pollen inside the
air or a fluid. The
roman Lucretius in his poem on the Nature of Things (60 BC) described
with amazing details the
motion of dust in the air, Jan Ingenhousz described the irregular motion
of coal dust on the
surface of alcohol in 1785, but Brownian motion is regarded as the discovery
of the botanist
Robert Brown in 1827, who observed pollen grains executing a jittery
motion in a fluid. Brown
initially thought
that the pollen particles were "alive", but repeating the experiment with
dust confirmed the idea
that the jittery motion of the pollen grains was due to the irregular
motion of the fluid particles.

The mathematics behind "Brownian motion" was first described by Thiele (1880),
 and then by Louis
Bachelier in 1900 in his PhD thesis on "the theory of speculation",
in which he presented a
stochastic analysis of the stock and option market. Albert Einstein's
independent research in
1905 brought to the attention of the physicists the main mathematical
concepts behind
Brownian motion and indirectly confirmed the existence of molecules and atoms
(at that time
the atomic nature of matter was still a controversial idea). As we will see
below, the mathematical
prototype behind Brownian motion became a very useful tool for the analysis
of many natural
phenomena.

Several articles and experiments followed Einstein's  and Marian
Smoluchowski's work and confirmed
that the molecules of water move randomly, therefore a small particle
suspended in the
fluid experiences a
random number
of impacts of random strength and direction in any short time. So, after
Brown's observations
of the irregular motion of "pollen grains executing a jittery motion", and
the idea of how to
interpret it as "the random motion of particles suspended inside the fluid",
the next step is to put all
this together in a firm mathematical model,
"the continuous time stochastic process."
The end result is a
convenient  prototype for many phenomena, and today's research on "Brownian
motion" is used widely
for the interpretation of many phenomena.

This tutorial is organized as follows: In Sec.\ 2, we give an introduction to
Brownian motion and classical random walk.
Sec.\ 3 presents different models for classical diffusion, the
Langevin equation, the approach through Fick's law, Einstein's approach,
the Fokker-Planck equation, and the central limit theorem.
In Sec.\ 4, the characteristics of anomalous diffusion are described,
and a typical example, the rotating annulus, is presented.
Sec.\ 5 introduces Continuous Time Random Walk, the waiting and the velocity
model are explained, methods to solve the equations are discussed, and
also the Levy distributions are introduced.
In Sec.\ 6, it is shown how, starting from random walk models, fractional
diffusion equations can be constructed.
In Sec.\ \ref{hamdiff} we show how a quasi-linear diffusion equation can be derived
for Hamiltonian systems.
Sec.\ 8 briefly comments on alternative ways to deal with anomalous diffusion,
Sec.\ 9 contains applications to physics and astrophysics, and Sec.\ 10
presents the conclusions.

\section{Brownian Motion and Random walks\label{bmrw}}

\subsection{Brownian Motion Interpreted as a Classical Random Walk
\label{Bmasrw}}

To build a firm base for the stochastic processes involved in Brownian motion,
we may start with a very simple example.

\begin{figure}[ht]
  \centering
  \includegraphics[width=8cm]{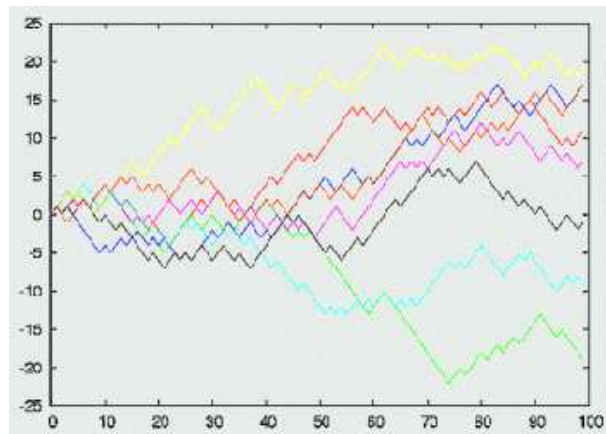}\\
  \caption{Random walk in one dimension (along the vertical axis)
as a function of time (to the right on the horizontal axis).}\label{rw1}
\end{figure}

We consider a random walk in one dimension (1D)
and assume that
the particles' steps $\D z$ are random and equally likely to either side,
left or right, and
of constant
length $\ell$ (see Fig. \ref{rw1}).
The position $z_N$ of a particle starting at $z_0=0$ after $N$ steps  is
\begin{equation}\label{steps1}
        z_N=\D z_N + \D z_{N-1} + ..... + \D z_1 = \sum_{i=1}^N \D z_i ,
\end{equation}
so that the squared length of the path equals
\begin{eqnarray}
z_N^2 &=&
\left(\sum_{j=1}^N \D z_j\right) \left(\sum_{k=1}^N \D z_k \right)
= \sum_{j,k=1}^N \D z_j \D z_k      \nonumber \\
&=& \sum_{j=1,\,k=j}^N \D z_j^2 + \sum_{j,k=1,\, k\ne j}^N \D z_j \D z_k
=N\ell^2+\sum_{{j,k=1}, \;{j\neq k}}^N \D z_j \D z_k .
\label{zsquare}
\end{eqnarray}
When averaging over a large number of particles, we find the mean squared
path length as
\begin{equation}\label{meanstep}
    <z_N^2>=N\ell^2 +  \left <\sum_{{j,k=1}, \;{j\neq k}}^N \D z_j \D z_k\right>.
\label{avzn}
\end{equation}
Each step of the walk is equally likely to the left or to the right, so that
the displacements $\D z_i$ are random variables with
zero mean. The products $\D z_j \D z_k$ are also random variables, and, since
we assume that $\D z_j$ and $\D z_k$ are independent of each other, the mean
value of the products is zero, so that
the expectation value of the mixed term in Eq.\ (\ref{avzn}) is zero. We thus find
\begin{equation}\label{meanstepaverage}
    <z^2>=N \ell^2 .
\end{equation}

The root-mean square displacement after $N$ steps of constant length $\ell$ (mean free path) is
\begin{equation}\label{meandstance}
 R:=   \sqrt{<z_N^2>}=\ell \sqrt{N}.
\end{equation}
We can now estimate the number of steps a photon starting from the Sun's core needs to reach the surface
of the Sun. From Eq.\ (\ref{meandstance}), we have
$N=(R/\ell)^2 $ and since the Sun's radius is $\sim 10^{10}\,$cm and the characteristic step (taken
into account the density in the solar interior) is $\sim 1\,$cm, we conclude that photons make $10^{20}$ steps
before exiting from the Sun's surface (this can answer questions like: if
the Sun's core stops producing energy, how long will it take until we feel the
difference on Earth?).

The mean free path $\ell$ can be estimated with a simple model. By assuming that a particle is moving inside a
gas with a mean speed $<v>,$ the distance traveled between two successive collisions is
$\ell=<v> \tau$, where $\tau$ is called collision time. If the particle has
radius $a$ and
travels a distance $L$ inside the gas with density $n$, then it will suffer $4\pi a^2 L n$ collisions, which
is just the number of particles in the volume $4\pi a^2 L $ the particle sweeps through.
The mean free path is then defined through the relation $4\pi a^2 \ell n=1$, i.e.\ $\ell$ is the
distance to travel and to make just one collision, so that
\begin{equation}\label{mfp}
    \ell=\frac{1}{4\pi a^2 n}.
\end{equation}
We may thus conclude that the number of steps a particle executes
inside a gas during a time $t$ is
$N=t/\tau$,
and, with Eq.\ (\ref{meanstepaverage}) and
the above relation $\ell=<v> \tau$,
the mean squared
distances it travels is
\begin{equation}\label{dif}
    <z^2>=N \ell^2=(t/\tau) (<v> \tau) \ell=(<v> \ell) t.
\end{equation}
Assuming that the random walk takes place in 3 dimensions
and that the gas is in equilibrium and isotropic, we expect that $<x^2>=<y^2>=<z^2>=\frac{<r^2>}{3}$, and
the mean square path length in 3 dimensions is
\begin{equation}\label{dif1}
    <r^2>=3<v> \ell t=Dt  ,
\end{equation}
where $D:=3 <v> \ell$ is called the diffusion coefficient,
which is a useful parameter to characterize particle diffusion
in the normal case (see Sec. \ref{diffusion}).
Important here is to note the linear scaling relation between $<r^2>$
and time $t$.

\subsection{Formal Description of the Classical Random Walk\label{formalcrw}}

More formally, we can define the classical random walk problem as follows.
We consider the position $\vec r$ of a particle in 1, 2, or 3  dimensional space,
and we assume that the position changes in repeated random steps $\Delta \vec r$.
The
time $\Delta t$ elapsing between two subsequent steps is assumed to be constant,
time plays thus a dummy role, it actually is a simple counter.
The position $\vec r_n$ of a particle after $n$ steps, corresponding to
time $t_n=n\Delta t$, is
\beq
\vec r_n = \Delta \vec r_n + \Delta \vec r_{n-1} + \Delta \vec r_{n-2} + \, ...
                                             \, + \Delta \vec r_1 + \vec r_0
\label{rnclass}
\eeq
where $\vec r_0$ is the initial position, and $\Delta \vec r_{i}$ is the $i$th
step (or increment, or displacement). The position $\vec r_n$ as well as the
increments $\Delta \vec r_{i}$ are all random variables. To specify the problem
completely, we have to prescribe the probability distribution
$q_{\Delta\vec r}(\Delta\vec r)$
for the increments $\Delta \vec r_{i}$, which
yields the probability for the particle to make a certain
step $\Delta \vec r_{i}$ (with given length and direction).
Writing $q_{\Delta\vec r}(\Delta\vec r)$ in this way, we have
made the assumptions that all the increments have the same probability
distribution and
that the increments are independent of each other (the value $\Delta \vec r_{i}$
takes in any realization is completely independent of the value
taken in the previous step by $\Delta \vec r_{i-1}$).
Generalizations to time-dependent increment distributions or correlated
increments are of course possible.

Since $\vec r_n$ is a random variable, the solution we are looking for in the random
walk problem is in the form of the probability distribution $P(\vec r,t_n)$, which yields
the probability for a particle to be at position $\vec r$ at time $t=t_n\equiv n\Delta t$.

If we are interested in the mean square displacement, we can just
square Eq.\ (\ref{rnclass}), and,
rearranging the terms in the same way as in Eq.\ (\ref{zsquare}), we find for $\vec r_0=0$
\beq
\langle \vec r_n^{\,2} \rangle
= \sum_{j=1,\,(k=j)}^n \langle \D \vec r_j^{\,2} \rangle
  + \sum_{j,k=1,\, k\ne j}^n \langle \D \vec r_j \D \vec r_k \rangle .
\label{vecrsquare}
\eeq
The first term on the right hand side is just a sum over the variances
$\sigma^2_{\Delta\vec r,\,j}$ of $q_{\Delta\vec r}(\Delta\vec r)$, since
by definition $\sigma^2_{\Delta\vec r} :=
\langle \D\vec r_j^{\,2} \rangle =
\int \Delta\vec r^{\, 2} q_{\Delta\vec r}(\Delta\vec r) \,d\Delta\vec r$ if the mean value of
$q_{\Delta\vec r}(\Delta\vec r)$ is zero. The second term on the
right hand side is the covariance $\textrm{cov}(\Delta\vec r_j,\Delta\vec r_k)$
of the random walk steps, and it is zero if the steps a particle takes
are independent of each other. We thus can write Eq.\ (\ref{vecrsquare})
as
\beq
\langle \vec r_n^{\,2} \rangle
= \sum_{j=1,\,(k=j)}^n  \sigma^2_{\Delta\vec r,\,j}
  + \sum_{j,k=1,\, k\ne j}^n \textrm{cov}(\Delta\vec r_j,\Delta\vec r_k).
\label{vecrsquare2}
\eeq
The particular random walk we considered in Sect.\ \ref{Bmasrw} 
can thus
be understood on the base of Eq.\ (\ref{vecrsquare2}) as a case with
zero covariance and variance $\sigma^2_{\Delta\vec r,\,j}=\ell^2$,
due to the constant step length,
which leads to the mean square displacement in Eq.\ (\ref{meanstepaverage}).

\section{Models for Normal Diffusion}\label{diffusion}

\subsection{Langevin's Equation\label{langeq}}

We turn now to a different way of treating Brownian motion.
We again consider a particle with mass $m$ 
performing a random walk inside a fluid due to the bombardment by
the fluid molecules, which obey an equilibrium  (Maxwellian)
distribution. Pierre Langevin \citep{Langevin08} described this motion with a simple but very interesting
stochastic differential equation (let us work in one dimension for simplicity),
\begin{equation}\label{langevin}
   m\ddot{x}=-a\dot{x}+F(t),
\end{equation}
where the term $a \dot{x}$ represents the friction force, $\dot{x}$ is the particle velocity,
$a$ is the damping rate and depends on the
radius of the particle and the viscosity of the fluid, and $F(t)$ is a random fluctuating force
due to the random bombardment of the particle by the fluid molecules.
If the random fluctuating force were absent, the particle starting with an initial velocity $v_0$ would
gradually slow down due to the friction term.
Multiplying Eq.\ (\ref{langevin}) with $x$, we have
\[
mx\ddot{x}=m\left[\frac{d (x \dot{x})}{dt}-\dot{x}^2\right]=-ax\dot{x}+xF(t) ,
\]
and after taking averages over a large number of particles we find, since
$<xF(t)>=0$ due to the irregular nature of the force $F(t)$,
\begin{equation}\label{langevin2}
        m\frac{d <x \dot{x}>}{dt}=m<\dot{x}^2>-a<x\dot{x}>    .
\end{equation}
Since the background gas is is in equilibrium, the kinetic energy of the particle is proportional to
the gas temperature,  $m<\dot{x}^2>/2=kT/2$, where $k$ is the Boltzmann constant and $T$ the temperature
of the gas. Eq. (\ref{langevin2}) now takes the form
\[
\left( \frac{d}{dt}+\gamma \right)<x\dot{x}>=\frac{kT}{m} ,
\]
where $\gamma =a/m$, which has the solution
\begin{equation}\label{langevin3}
        <x\dot{x}>=\frac{1}{2}\frac{d <x^2>}{dt}=Ce^{-\gamma t}+\frac{kT}{a}.
\end{equation}
At $t=0$,  the mean square displacement is zero, so that $0=C+kT/a$,
and Eq.\ (\ref{langevin3}) becomes
\[
 \frac{1}{2}\frac{d <x^2>}{dt}=\frac{kT}{a}(1-e^{-\gamma t}).
\]
On integrating the above equation we find the solution
\begin{equation}\label{langevin4}
        <x^2>=\frac{2kT}{a}\left [t-\frac{1}{\gamma} \left(1-e^{-\gamma t}\right)\right].
\end{equation}
In the limit $t<<1/\gamma$ (time much shorter than the collision time) the
solution in Eq. (\ref{langevin4}) is of the form $<x^2>\sim t^2$ (expanding the
exponential up to second order), which is called "ballistic" diffusion and
means that at small times particles are not hindered by collisions yet and
diffuse very fast, see Sect.\ \ref{anomdef}.
In the other limit, $t>>1/\gamma$, the
solution has the form
\begin{equation}\label{normal1}
        <x^2>\sim \frac{2kT }{a}t ,
\end{equation}
or, for the 3 dimensional case, if again the gas is in
equilibrium and isotropic so that $<r^2>/3=<x^2>,$
\begin{equation}\label{normal}
        <r^2>=\frac{6kT }{a}t=Dt ,
\end{equation}
where $D=6kT/a$ is an expression for the diffusion
constant in terms of particle and fluid characteristics
(cf.\ Eq. (\ref{dif1})), and note that again $<r^2>$ has a simple
scaling relation with time, $<r^2>=Dt$, as in Sect.\ \ref{Bmasrw}. 

\subsection{Modeling Diffusion with Fick's Law\label{classappr}}

Diffusion usually occurs if there is a spatial difference in concentration
of e.g.\ particles or heat etc., and it usually acts such as to reduce the
spatial inhomogeneities in concentration.

Let us consider particle diffusion along the $z$-direction in 3-dimensional
space, and let us assume
that two  elementary areas perpendicular to the flow
(in the $x$-$y$-plane) are a distance $\Delta z$ apart.
Particle conservation implies that
the time variation of the density $n(z,t)$ inside
the elementary volume $\Delta x\Delta y\Delta z$ equals the inflow
minus the outflow of particles, so that, if $J(z,t)$ denotes the
particle flux,
\[
\frac{\partial n(z,t)}{\partial t}  \Delta x \Delta y \Delta z
=
J(z)\Delta x \Delta y -J(z+\Delta z) \Delta x \Delta y
=
-\frac{\partial J}{\partial z} \Delta x \Delta y \Delta z
\]
which leads to the diffusion equation in its general form,
\beq
    \frac{\partial n(z,t)}{\partial t}= -\frac{\partial J(z,t)}{\partial z}.
\label{diffusioneq0}
\eeq

The problem that remains is to determine the particle flux $J$. From its
physical meaning, it obviously holds that
\beq
J(z,t) = n(z,t) v(z,t)  ,
\eeq
where $v(z,t)$ is an average particle flow velocity. Using this expression
in Eq.\ (\ref{diffusioneq0}) leads to a closure problem, we would need to find
ways to determine $v(z,t)$.

It is well documented experimentally that the flux of particles $J$ crossing
a certain area (again, say in the $x$-$y$-plain) is proportional to the
density gradient along the $z$-axis (Fick's Law),
\begin{equation}\label{fick}
        J_z=-D(z)\frac{\partial n}{\partial z} ,
\end{equation}
where $D$ is the diffusion coefficient discussed already in the previous
sections, and which generally may also depend on $z$.
With Eq.\ (\ref{fick}),
the diffusion equation takes the classical form
\begin{equation}\label{diffusioneq}
    \frac{\partial n(z,t)}{\partial t}=\frac{\partial }{\partial z}D(z)\frac{\partial n(z,t)}{\partial z}  ,
\end{equation}
or, for constant diffusion coefficient,
\begin{equation}\label{diffusioneq1}
    \frac{\partial n(z,t)}{\partial t}=D\frac{\partial ^2 n(z,t)}{\partial z^2}.
\end{equation}

In infinite space, and if all particles start initially from $z=0$,
the solution of Eq. (\ref{diffusioneq1}) is
\begin{equation}\label{solDiff}
    n(z,t)=\frac{N_0}{\sqrt{4 \pi D t}}e^{-z^2/4Dt}  ,
\label{nztG}
\end{equation}
where $N_0$ is the total number of particles inside the volume under
consideration. The solution obviously is identical to a Gaussian distribution
with mean zero and variance $2Dt$. The variance is defined as
\begin{equation}\label{normalDiff}
    <z^2(t)>=\int z^2 n(z,t) \, dz=2Dt ,
\end{equation}
which is just identical to the mean square displacement, so that
the results obtained earlier, using the simple version of
the random walk in Sect.\ \ref{Bmasrw} 
or the stochastic differential equation of Langevin in Sect.\ \ref{langeq},
are again confirmed.

Diffusion obeying Eq.\ (\ref{normalDiff}) is called \textbf{normal diffusion}
and is characteristic for the diffusion processes in systems that
are in equilibrium or very close to equilibrium.
{\bf Generalizing the
above results (Eqs. (\ref{diffusioneq}), (\ref{diffusioneq1}), (\ref{solDiff}),
(\ref{normalDiff}) is
simple and can be found in the literature (see references).}

\subsection{Einstein's Formalism for the Classical Random Walk and the
Diffusion Equation\label{einsform}}

A different approach to treat normal diffusion was introduced
by Bachelier and by Einstein \citep[see][]{Einstein05}.
Here, the starting point is the classical random walk
as defined in Sec.\ \ref{formalcrw}, and we consider the 1-dimensional case.
According to Sec.\ \ref{formalcrw}, the solution of the random walk
problem is in the form of the probability distribution $P(z,t)$ for
a particle at time $t$ to be at position $z$. Assume that we would
know the distribution $P(z,t-\Delta t)$ one time-step $\Delta t$ earlier
(remember $\Delta t$ is assumed constant). If particles are conserved,
the relation
\beq
P(z,t) = P(z-\Delta z,t-\Delta t) \, q_{\Delta z}(\Delta z).
\label{eins1}
\eeq
must hold, with $q_{\Delta z}$ the distribution of random walk steps.
Eq.\ (\ref{eins1}) states that the probability to be at time $t$ at position
$z$ equals the probability to have been at position $z-\Delta z$ at time
$t-\Delta t$, and to have made a step of length $\Delta z$. We still have
to sum over all possible $\Delta z$, which leads to the Einstein (or Bachelier)
diffusion equation,
\beq
 P(z,t) = \int_\infty^\infty P(z-\Delta z,t-\Delta t) \,q_{\Delta z}(\Delta z)\,d\Delta z
\label{eins2}
\eeq
This is an integral equation that determines the solution $P(z,t)$ of the
random walk problem as defined in Sec.\ \ref{formalcrw}. The power of this
equation will become clear below when we will show ways to treat cases of
anomalous diffusion. Here, we still focus on normal diffusion. As will become
clear later, it is actually a characteristic of normal diffusion that the
particles take only small steps $\Delta z$ compared to the system size.
This implies that
$q_{\Delta z}(\Delta z)$ is non-zero only for small $\Delta z$,
the integral in Eq.\ (\ref{eins2}) is only over a small $\D z$-range, and we
can expand $P(z-\Delta z,t-\Delta t)$ in $z$ and $t$ (also
$\Delta t$ is small),
\bea
P(z-\Delta z,t-\Delta t) &=& P(z,t) - \Delta t\,\partial_t P(z,t) - \Delta z\,\partial_z P(z,t)\nonumber\\
&+& \frac{1}{2} \Delta z^2 \, \partial_z^2 P(z,t)
\label{expP}
\eea
Inserting into Eq.\ (\ref{eins2}), we find
\bea
P(z,t) &=& \int P(z,t)\,q_{\Delta z}(\Delta z)\,d\Delta z
- \int \Delta t\partial_t P(z,t)\,q_{\Delta z}(\Delta z)\,d\Delta z \nonumber \\
&-& \int \Delta z\partial_z P(z,t)\,q_{\Delta z}(\Delta z)\,d\Delta z\nonumber \\
&+& \int \frac{1}{2} \Delta z^2\,  \partial_z^2 P(z,t)\,q_{\Delta z}(\Delta z) \,d\Delta z
\eea
Obviously, $P(z,t)$, its derivatives, and $\Delta t$ are not affected by the integration, furthermore we use the normalization of $q_{\Delta z}$
($\int q_{\Delta z}(\Delta z)\,d\Delta z = 1$), we assume $q_{\Delta z}$
to be symmetric (mean
value zero, $\int \Delta z \,q_{\Delta z}(\Delta z)\,d\Delta z = 0$), and we use
the definition of the variance $\sigma^2_{\Delta z}$
($\int \Delta z^2 \,q_{\Delta z}(\Delta z)\,d\Delta z = \sigma^2_{\Delta z}$), so that
\beq
P(z,t) = P(z,t) -\Delta t \,\partial_t P(z,t)
+ \frac{1}{2} \sigma^2_{\Delta z} \,\partial^2_z P(z,t)
\eeq
or
\beq
\partial_t P{z,t} = \frac{\sigma^2_{\Delta z}}{2\Delta t} \partial_z^2 P(z,t)
\label{einsdiff}
\eeq
i.e.\ we again recover the simple diffusion equation, as in Sec.\ \ref{classappr},
with diffusion coefficient $D=\frac{\sigma^2_{\Delta z}}{2\Delta t}$,
and the solution to it in infinite space is again the Gaussian of Eq.\ (\ref{nztG}).

\subsection{Fokker-Planck Equation\label{fokkpla}}


The Fokker-Planck (FP) equation (or Kolmogorov forward equation)
is a more general diffusion equation than the simple
equations introduced in
Secs.\ \ref{classappr} an \ref{einsform}.
We again start from a description of diffusion in terms of a random
walk, as in Sec.\ \ref{einsform},
we though relax two assumptions made there: (i) We assume now
that the mean value $\mu_{\D z}$ of the random walk steps
can be different from zero,
which corresponds to a systematic motion of the particles
in the direction of the sign of $\mu_{\D z}$, and (ii) we assume that
both the mean and the variance can be spatially dependent,
$\mu_{\D z}=\mu_{\D z}(z)$ and
$\sigma_{\D z}^2=\sigma_{\D z}^2(z)$, which means
that the distribution of increments depends on the spatial location,
i.e.\ it is of the form $q_{\Delta z,z}(\Delta z,z)$. To be compatible
with these
assumptions, Eq.\ (\ref{eins2}) must be rewritten in a slightly
more general form,
\beq
P(z,t) = \int_\infty^\infty P(z-\D z,t-\D t)
q_{\D z,z}(\D z,z-\D z) \,d\D z  ,
\label{chapkol}
\eeq
which is the {\it Chapman-Kolmogorov equation}, and where
now $q_{\D z, z}(\D z,z)$ is the probability
density for being at position $z$ and making a step $\D z$
in time $\D t$.
The FP equation
can be derived in a way similar to the one presented
in Sec.\ \ref{einsform}:
We expand the integrand of Eq.\ (\ref{chapkol})
in a Taylor-series in terms of $z$, so that
$P(z,t) = \int_\infty^\infty A B \,d\D z$,
with
\beq
 A=P(z,t)-\p_t P(z,t)\D t-\p_z P(z,t)\D z+ \frac{1}{2}\p_z^2 P(z,t)\D z^2 + ... ,
\label{Aeq}
\eeq
where we have also expanded to first order in $t$, and
which is of course the same as Eq.\ (\ref{expP}),
and newly we have
\beq
B=q_{\D z,z}(\D z,z)-\p_z q_{\D z,z}(\D z,z)\D z+ \frac{1}{2}\p_z^2
q_{\D z,z}(\D z,z)\D z^2 + ...
\label{Beq}
\eeq
(note that the Taylor expansion is with respect to the second
argument of $q_{\D z,z}$, we expand only with respect to $z$,
not though with respect to $\D z$).
In multiplying and evaluating the integrals, we use the normalization
of $q_{\D z,z}$ ($\int q_{\D z,z}(\D z,z)\,d\D z)=1$),
the definition of
the mean value \\
($\mu_{\D z}(z) :=  \int \D z q_{\D z,z}(\D z,z)\,d\D z$) and of the
second moment
($\langle\D z^2\rangle (z) := \int \D z^2 q_{\D z,z}(\D z,z)\,d\D z$),
and expressions
like $\int \D z \, \p_z q_{\D z,z}(\D z,z)\,d\D z $ are considered to equal
$\p_z \int \D z  q_{\D z,z}(\D z,z)\,d\D z\equiv \p_z \mu_{\D z}(z)$, so that,
on keeping all terms up to second order in $\D z$, we find
the {\it Fokker-Planck equation},
\beq
\p_t P(z,t) = -\p_z[V(z)P(z,t)]+\p_z^2 [D(z)P(z,t)] ,
\label{FokkerP}
\eeq
with 
$V(z)\equiv\mu_{\D z}(z)/\Delta t$ a drift velocity,
and $D(z)\equiv \langle\D z^2\rangle (z)/2\Delta t$
the diffusion coefficient
\citep[for a 3 dimensional formulation see e.g.][]{Gardiner04}.
The basic difference between the FP equation and the
simple diffusion equation in Eq.\ (\ref{einsdiff}) is the appearance
of a drift term, and that both the drift velocity and the diffusion
coefficient are allowed to be spatially dependent
(Fick's law also allows a spatially dependent diffusion coefficient,
see Eq.\ (\ref{diffusioneq})).
These differences allow the FP equation to model
more complex diffusive behaviour.

The FP equation is also applied to velocity space, e.g.\
in plasma physics in order to treat
collisional effects, or to position and velocity space together.
It has the advantage of being a deterministic differential equations
that allows to describe the evolution of stochastic systems, as long as
the diffusivities and drift velocities are known, and as long as the
conditions for its applicability are met, see the remarks below.

We can illustrate the typical use of the FP equation on the example of
Brownian motion in the Langevin formalism in Sec.\ \ref{langeq},
which allowed us to calculate the diffusion coefficient in
Eq.\ (\ref{normal}). If we are interested in the evolution of the
distribution of particles $P$, then with the Langevin formalism
we would have to follow a large number of individual particles
over the times of interest and then to construct the distribution
function, which may become very intense in computing
effort. Instead, one can use the diffusivity from Eq.\ (\ref{normal}),
insert it into the FP equation, and solve the FP equation.
Since in this example the diffusion coefficient is constant and there
is no drift velocity, the FP equation reduces to the
simple diffusion equation (\ref{diffusioneq1}), whose solution for
$P$ is given in Eq.\ (\ref{solDiff}).

We just note that in the general case where the diffusion coefficient
is $z$-dependent, $D=D(z)$, there is a small
difference between the diffusive term in the Fokker-Planck equation
and the Fickian diffusion equation, Eq.\ (\ref{diffusioneq}), in that
the diffusivity is only once differentiated in the latter. This
difference and its consequences are analyzed in \citet{Milligen05}.

>From its derivation it is clear that the FP
equation is suited only for systems close to equilibrium,
with just small deviations of some particles from equilibrium, or, in the
random walk sense, with just small steps of the particles performing the
random walk, exactly as it holds for the simple diffusion equation
in Sect.\ \ref{einsform}.

A further natural generalization for a diffusion
equation in the approach followed here would be
not to stop the Taylor expansion in Eqs.\ (\ref{Aeq}) and (\ref{Beq})
at second order in $z$,
but to keep all terms, which would lead to the so-called
{\it Kramers-Moyal expansion}.

More details about the Fokker-Planck equation can be found in the
literature \citep[e.g.][]{Gardiner04}.


\subsection{Why Normal Diffusion Should Be the Usual Case\label{CLT}}

The appearance of normal diffusion in many natural phenomena close
to equilibrium
and the particular Gaussian form of
the solution of the diffusion equation
can also be understood from
probability theory. The Central Limit Theorem (CLT) states that if a statistical
quantity (random variable) is the sum of many other statistical quantities,
such as the position of a random walker after $n$ steps according to
Eq.\ (\ref{rnclass}), and if (i) all the $\D z_i$ have finite mean $\mu_{\D z}$
and variance $\sigma_{\D z}^2$, (ii) all the $\D z_i$ are mutually independent, and
(iii) the number $n$ of the $\D z$ is large, then, independent of the
distribution of the $\D z_i$'s, the distribution $P(z,t_n)$ of $z_n$ is a Gaussian. In
particular, if $\mu_{\D z}= 0$, $z_0=0$ and all the $\D z_i$ have the same variance, then
\beq
P(z,t_n) = \frac{1}{2\pi n \sigma_{\D z}^2} e^{-\frac{z^2}{2n\sigma_{\D z}^2}}
\label{CLTG}
\eeq
with variance $\sigma_{z_n}^2 = n\sigma_{\D z}^2$, or, if we set $n=t/\D t$, then
$\sigma_{z_n}^2 = t_n\sigma_{\D z}^2/\D t$.

The mean
square displacement equals per definition the variance,
\beq
\langle z(t_n)^2\rangle = \int z^2 P(z,t_n) \,dz=\sigma_{z_n}^2 = t_n\sigma_{\D z}^2/\D t
\eeq
(for $z_0=0$), and diffusion is thus always normal in the cases where the CLT
applies.

Moreover, the CLT predicts quantities that are the result
of many small scale interactions to be distributed according
to a Gaussian, and indeed this is what we found for the
distribution of the classical random walker, see Eq.\ (\ref{solDiff}).
Stated differently, we may say that
the appearance of non-Gaussian distributions
is something unexpected and unusual according to the CLT.
We just mention
that also the equilibrium velocity distributions of gas or fluid particles
are in accordance with the CLT, the velocity components,
say $v_x$, $v_y$, $v_z$, follow Gaussian distributions,
and therewith the magnitude
$v=\sqrt{v_x^2+v_y^2+v_z^2}$ exhibits a Maxwellian distribution. Again
then, the appearance of non-Maxwellian velocity distributions is unexpected
on the base of the CLT.

\section{Anomalous Diffusion\label{anodi}}

\subsection{Systems Far from Equilibrium: The rotating Annulus}

\begin{figure}[ht]
\centering
  \includegraphics[width=8cm]{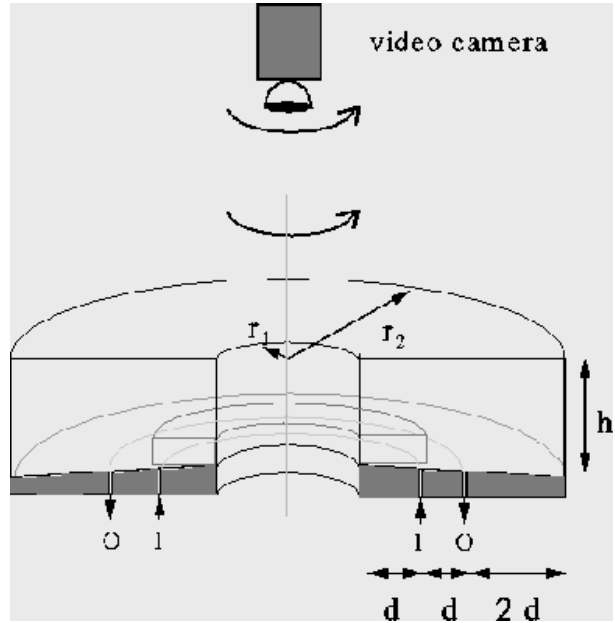}\\
  \caption{Rotating annulus.}\label{ra}
\end{figure}

\begin{figure}[ht]
\centering
  \includegraphics[width=5cm]{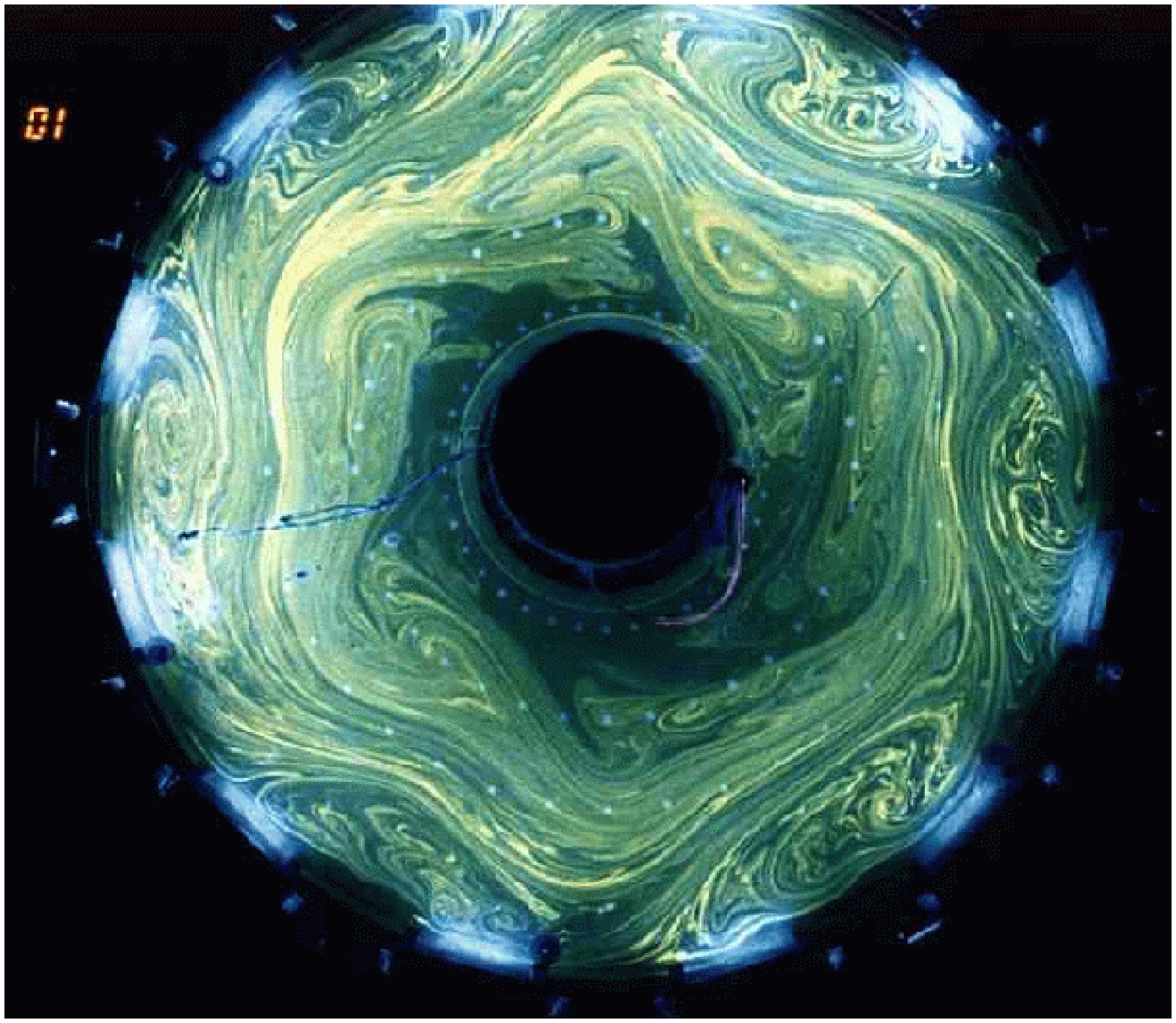}
  \includegraphics[width=5cm]{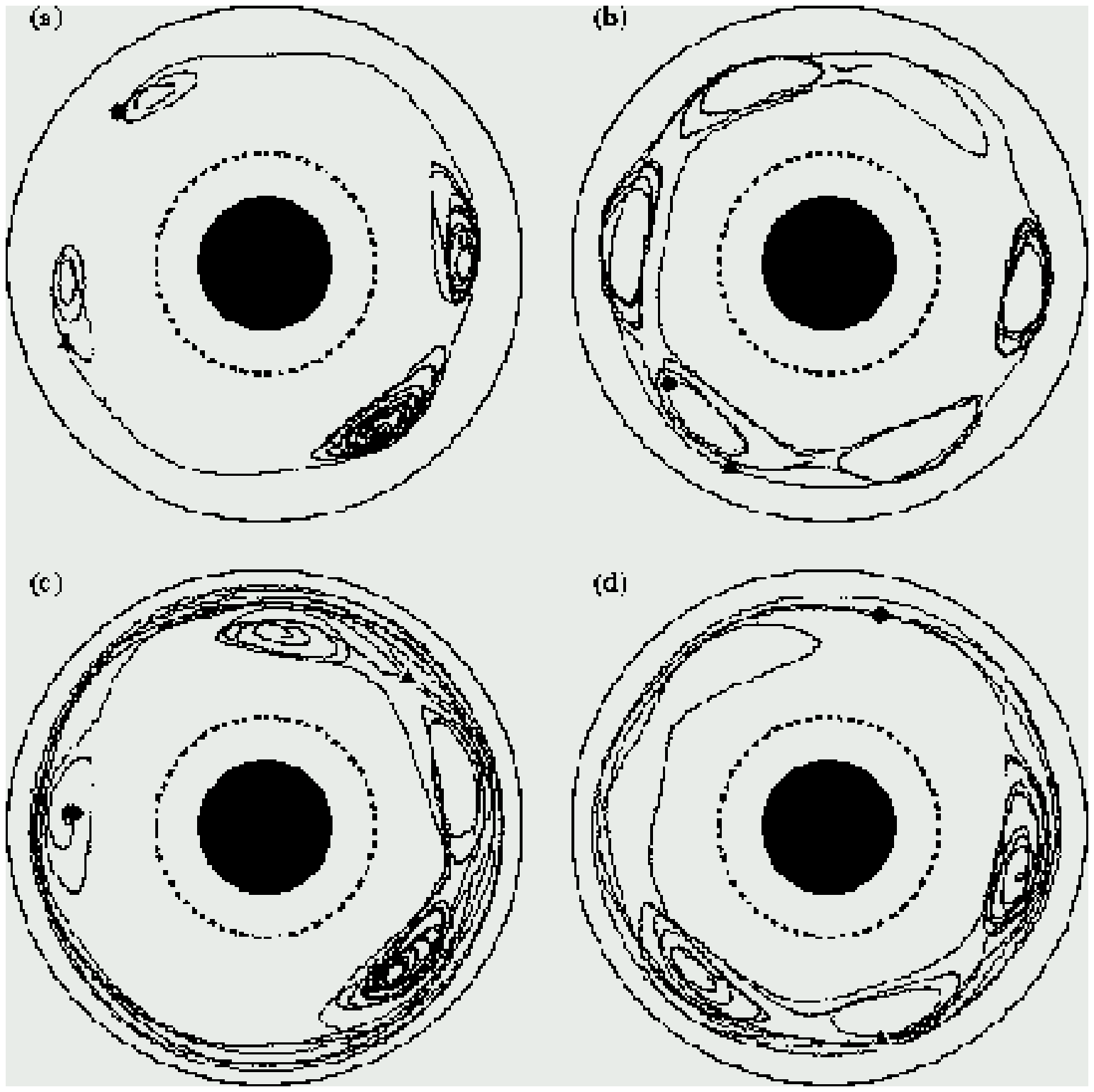}\\
  \caption{(a) The formation of eddies inside the rotating annulus, as
recorded by the camera (left panel), and (b) typical orbits of tracer particles
inside the annulus (right panel).}\label{anorbits}
\end{figure}

The simple experiment of the rotating annulus, shown in Fig. \ref{ra},
allows to illustrate
the differences between normal and
anomalous diffusion \citep{Solomon94,Weeks96}.
Water is pumped into the annulus through a ring of holes marked with $I$
and pumped out through a second ring of holes marked with $O$.
The annulus is completely
filled with water and rotates as a rigid body (the inner and outer walls
rotate together). The pumping of the fluid generates a turbulent
flow in the annulus.
A camera on top of the annulus
records the formation of the turbulent eddies inside the rotating annulus and
allows to track seeds of
different tracer particles injected into the fluid and to
monitor their orbits
(see Fig. \ref{anorbits}).

In the case of normal diffusion, which
occurs mainly in fluids close to equilibrium,
the particle trajectories are characterized by irregular,
but small steps, which makes trajectories look
irregular but still homogeneous (see Fig.\ \ref{rw1}).
The trajectories shown
in Fig.\ \ref{anorbits} for the highly turbulent rotating annulus,
which is far away from equilibrium,
show different types of orbits, with two basic new characteristic,
there is ``trapping" of particles
inside the eddies, where particles stay for 'unusually' long times
in a relatively small spatial area,
and there are "long flights" of particles, where particles
are carried in one
step over large distances, in some cases almost through the entire system.


\subsection{The Scaling of "Anomalous" Trajectories\label{anomdef}}

Normal diffusion
has as
basic characteristic the linear scaling of the mean square displacement
of the particles with time, $\langle r^2 \rangle\sim Dt$.
Many different experiments though, including the one shown in the previous
section, reveal
deviations from normal diffusion, in that diffusion is either
faster or slower, and which is termed anomalous diffusion. A useful
characterization of the diffusion process is again through the scaling
of the mean square displacement with time, where though now we are
looking for a more general scaling of the form
\beq
\langle r^2(t) \rangle \sim t^\gamma  .
\label{asca}
\eeq
Diffusion is then classified through the scaling index $\gamma$. The
case $\gamma=1$ is normal diffusion, all other cases
are termed anomalous. The cases $\gamma>1$ form the family of super-diffusive
processes, including the particular case $\gamma=2$, which is called ballistic
diffusion, and the cases $\gamma<1$ are the sub-diffusive processes. If the
trajectories of a sufficient number of particles inside a 
system are known, then plotting $\log <r^2>$ vs $\log t$ is an experimental
way to determine the type of diffusion occurring in a given
system.

\begin{figure}[ht]
\centering
  \includegraphics[width=5cm]{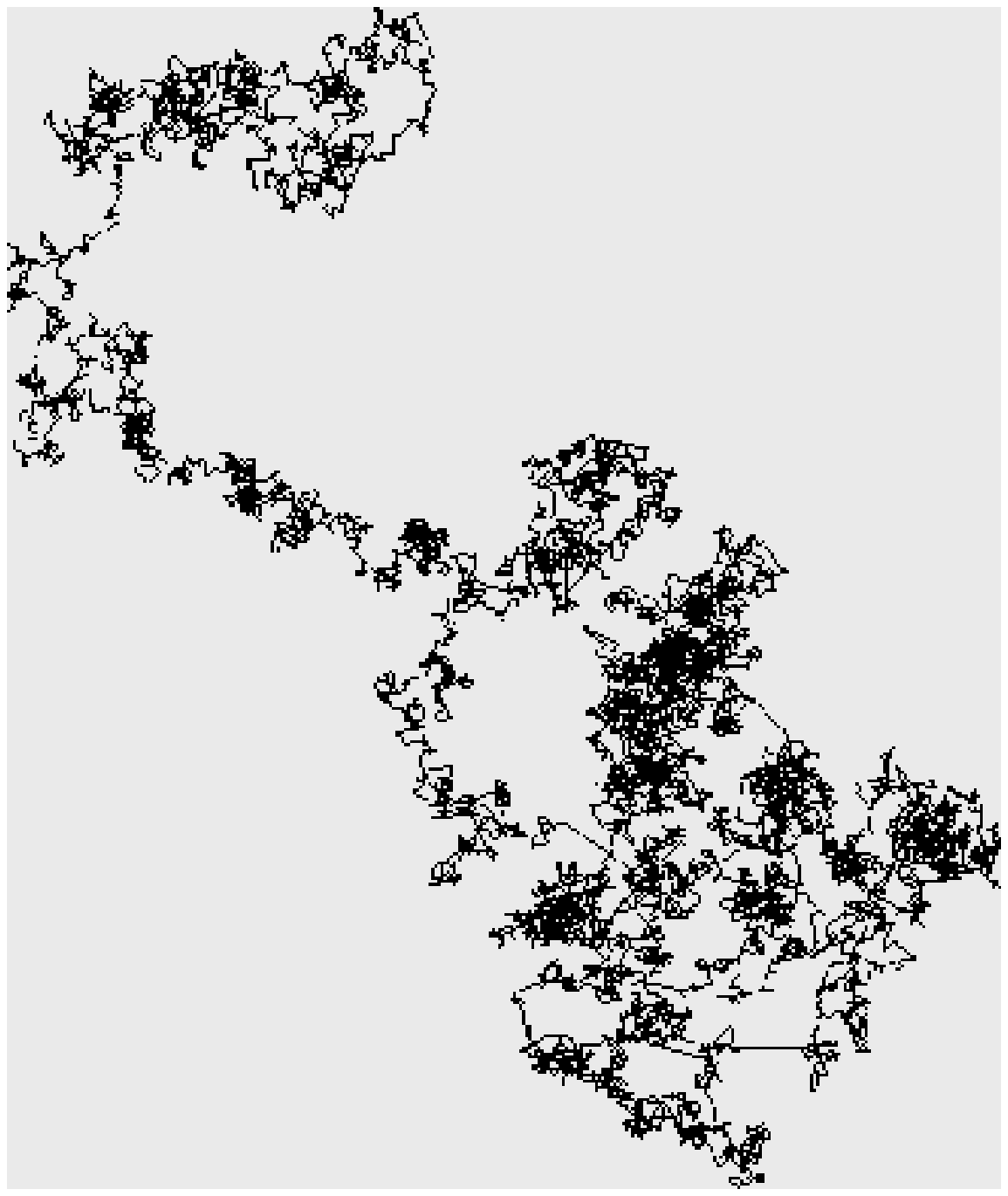}
  \includegraphics[width=5cm]{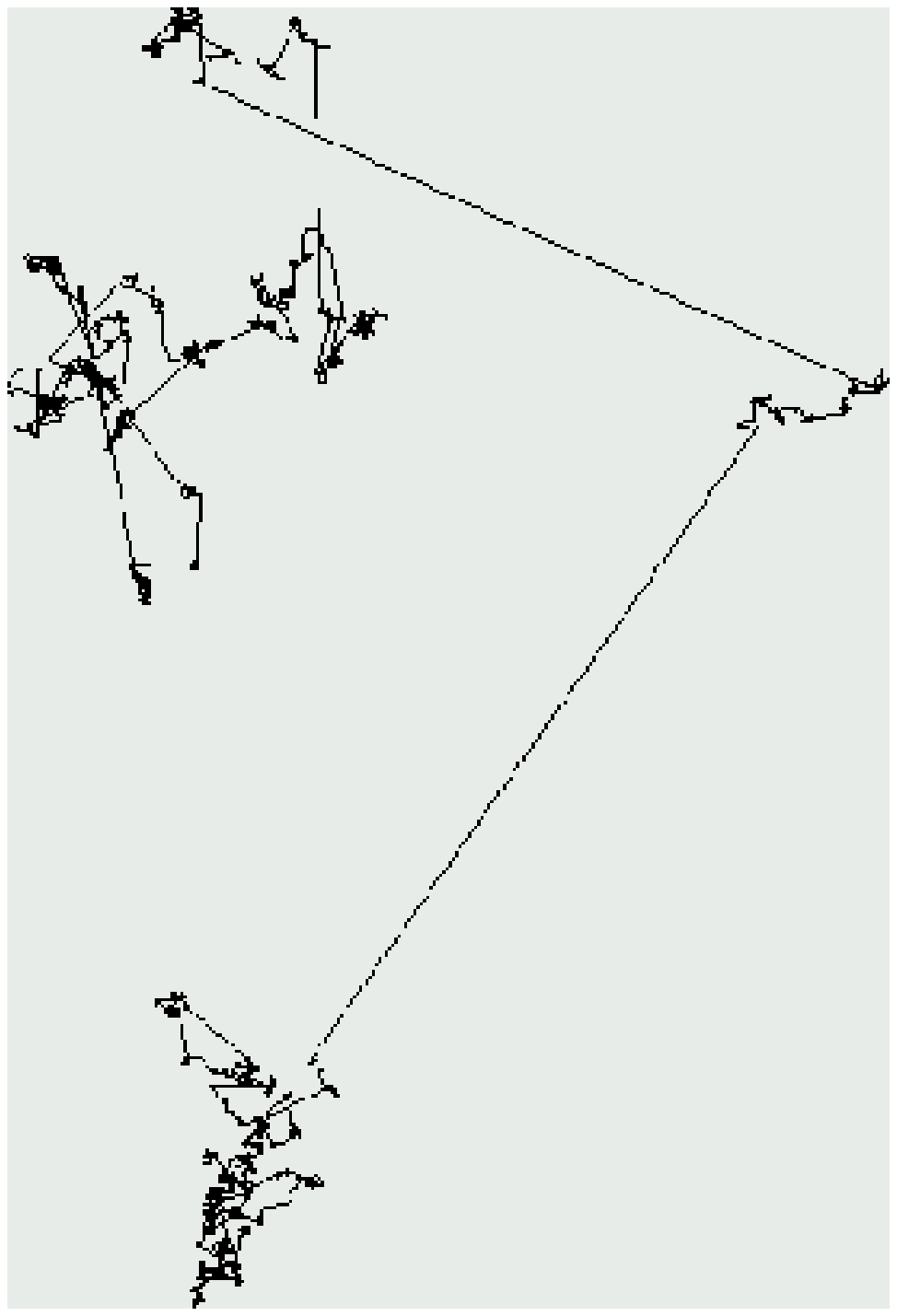}\\
  \caption{(a) Random walk in dynamical systems close to equilibrium
(normal diffusion; trajectory on the left), (b) random walk in dynamical
systems far from equilibrium (anomalous diffusion; trajectory on the right).}
\label{levy}
\end{figure}

As an illustration, let us consider a particle that is moving with
constant velocity $v$ and undergoes no collisions and experiences no
friction forces. It then
obviously holds that $r=vt$, so that $\langle r^2(t)\rangle\sim t^2$.
Free particles are thus super-diffusive in the terminology used here,
which is also the origin of the name ballistic for the case $\gamma=2$.
Accelerated particles would even diffuse faster. The difference between
normal and a anomalous diffusion is also illustrated in Fig. \ref{levy},
where in the case of anomalous diffusion
long "flights" are followed by efficient "trapping" of particles in
localized spatial regions,
in contrast to the more homogeneous
picture of normal diffusion.

It is to note that anomalous diffusion manifests itself not only
in the scaling of Eq.\ (\ref{asca}) with $\gamma\ne 1$ (which
experimentally may also be difficult to be measured), but also in 'strange'
and 'anomalous' phenomena such as 'uphill' diffusion, where particles
or heat diffuse in the direction of higher concentration, or the appearance
of non-Maxwellian distributed particle velocities (see Sec.\ \ref{CLT}),
very often of power-law
shape, which is very common in high energy astrophysics (e.g.\ cosmic rays),
etc.

\section{Continuous Time Random Walk}

\subsection{Definition\label{subsub1}}

Given the experimental ubiquity of anomalous diffusion phenomena, the
question arises of how to model such phenomena. One way of tackling it
is through the random walk formalism. So-far, we have used the random
walk to model classical diffusion, and in Sec.\ \ref{einsform} it had
been shown how the random walk is related to a simple diffusion equation
if the steps the particles take on their walk are small. One way to model
anomalous diffusion is by relaxing the latter condition, and to allow the
particles to also take large steps, where 'large' in a finite system means
large up to system size, and in infinite systems it means that the steps
are unbounded in length. Useful in this context is the family of Levy
distributions as step-size distributions $q_{\Delta z}$. They are
defined in closed form in Fourier space (see Sec.\ \ref{symmLevy} below),
and they have the property that
\beq
q^{L,\alpha}_{\Delta z}(\Delta z) \sim \vert \Delta z\vert^{-1-\alpha},
\ \ \ \textrm{for}\ \,\vert\Delta z\vert\ \,\textrm{large},\ \ \, 0 < \alpha < 2,
\eeq
so that there is always a small, though finite probability for any arbitrarily large
steps size. The Levy distributions all have an infinite variance,
$\sigma_{L,\alpha}^2=\int\Delta z^2 q^{L,\alpha}_{\Delta z}(\Delta z) \,d\Delta z=\infty$,
which makes their direct use as a step size distribution in the
classical random walk of Sec.\ \ref{formalcrw} and Eq.\ (\ref{rnclass})
impossible:
Consider the case of a random walk in 1-D, with the position of the
random walker after $n$ steps given by the 1-D version of Eq.\ (\ref{rnclass}),
and the mean square displacement (for $z_0=0$)
given by Eq.\ (\ref{vecrsquare2}).
Let us assume that
the steps are independent of each other, so that the covariances are zero
and the mean
square displacement is
$\langle z_n^2\rangle = n \sigma_{L,\alpha}^2$,
which is infinite, already after the first step.

A way out of the problem is to release time from its dummy role and make
it a variable that evolves dynamically, as the walker's position does.
In this way,
infinite steps in space can be accompanied by an infinite time for
the step to be completed, and the variance of the random walk, i.e.\ its
mean square displacement, remains finite. The extension of the random
walk to include the timing is called Continuous Time Random Walk (CTRW).
Its formal definition consists again of Eq.\ (\ref{rnclass}), as described
in Sec. \ref{formalcrw}, and, moreover, the time at which the $n$th step of
the walk takes
place is now also random (a random variable), and it evolves according
to
\beq
t_n = \Delta t_n + \Delta t_{n-1} + \Delta t_{n-2} + \, ...
                                             \, + \Delta t_1 + t_0,
\label{tctrw}
\eeq
where $t_0$ is the initial time, and the $\Delta t_i$ are random
temporal increments. To complete the definition of the CTRW, we need
also to give the probability distribution of the $\Delta t_i$, i.e.\ we
must specify the probability for the $i$th step to last a time $\Delta t_i$.

Two case are usually considered (not least to keep the technical problems
at a manageable level). (i) In the waiting model, the steps in position
and time are independent, and one specifies two probabilities,
one for $\D\vec r$, the $q_{\D\vec r}$ already introduced, and one
for $\D t$, say $q_{\D t}$. Here then $\D t$ is interpreted as a waiting time,
the particle waits at its current position until the time $\D t$ is elapsed,
and then it performs a spatial step $\D \vec r$ during which no time is
consumed
\citep[e.g.][]{Montroll65}.
(ii) In the velocity model, the time $\D t$ is interpreted as the traveling
time of the particle, $\D t=\vert \D \vec r\vert /v$, where $v$ is an
assumed constant
velocity (the velocity dynamics is not included, usually, see though
Sec.\ \ref{velo}), so that
the distribution of increments is
$q_{\D z,\D t} = \delta(\D t - \vert \D \vec r\vert/v)q_{\D \vec r}(\D \vec r)$
\citep[e.g.][]{Shlesinger87}.
We just note that in the general case one would have to specify the joint
probability distribution $q_{\D z,\D t}(\D z,\D t)$ for the spatial and
temporal increments.

\subsection{The CTRW Equations}

The CTRW equations can be understood as a generalization of the Einstein
equation, Eq.\ (\ref{eins2}), or the Chapman-Kolmogorov equation,
Eq.\ (\ref{chapkol}). It is useful to introduce the concept of the
turning-points, which are the points at which a particle arrives at and starts
a new random walk step. The evolution equation of the distribution of turning points
$Q(z,t)$ (here in 1-D) follows basically from particle conservation,
\bea
Q(z,t) &=& \int d\D z\int_0^t d\D t Q(z-\D z,t-\D t) q_{\D z,\D t}(\D z,\D t)
\nonumber\\
&& +\, \delta(t) P(z,t=0) + S(z,t),
\label{Qeq}
\eea
where the first term on the right side describes a completed random walk step, including
stepping in space and in time, the second term takes the initial condition
$P(z,t=0)$ into account, and the third term $S$ is a source term
\citep[see e.g.][]{Zumofen93}.

The expression for $P(z,t)$, the probability for the walker to be at position $z$ at time $t$,
is different for the waiting
and for the velocity model, respectively.
In case of the waiting model, where
$q_{\D z,\D t}(\D z,\D t)=q_{\D t}(\D t)q_{\D z}(\D z)$, we have
\beq
P_W(z,t) = \int_0^t d\D t\, Q(z,t-\D t) \Phi_W(\D t)   ,
\label{PWeq}
\eeq
with $\Phi_W(\D t):=\int_{\D t}^\infty dt^\prime q_{\D t}(t^\prime)$ the probability
to wait at least a time $\Delta t$ \citep[e.g.][]{Zumofen93}.

In the velocity model, where
$q_{\D z,\D t}(\D z,\D t) =
\delta(\D t - \vert \D z\vert/v) \,q_{\D z}(\D z)$,
$P(z,t)$ takes the form
\beq
P_V(z,t) = \int_{-vt}^{vt} d\D z \int_0^t d\D t\, Q(z-\D z,t-\D t) ,
\Phi_V(\D z,\D t)
\label{PVeq}
\eeq
with
\beq
\Phi_V(\D z,\D t) = \frac{1}{2} \delta(\vert \D z\vert -v \D t)
\int_{\vert\D z\vert}^\infty dz^\prime
\int_{\D t}^\infty dt^\prime
\delta(t^\prime-\vert z^\prime\vert/v) q_{\D z}(z^\prime)
\eeq
the probability to make a step of length at least $\vert \D z\vert$ and of
duration at least $\D t$ \citep[e.g.][]{Zumofen93,Shlesinger87}.

Both, the expression for $P_W$ and $P_V$ determine the probability for
seeing the particle when moving in-between two turning points, taking into
account only the part of the random walk in which time is consumed by the particle.

The kind of diffusion that the CTRW formalism yields depends on the distribution
of step increments. If the increments are small, then the treatment of Sec.\ \ref{einsform}
can be applied again, diffusion is normal, and again a simple diffusion equation
can be derived. If the increments are not small, then super- as well as sub-diffusion
can result, depending on the concrete choice of increment distributions. For instance,
small spatial steps in combination with Levy-distributed, long waiting times will
yield sub-diffusion in the waiting model. An important property of the CTRW equations
is that they are non-local, both in space and time (which is also termed non-Markovian).
Anomalous diffusion phenomena in the CTRW approach are thus considered non-local
processes, and with that they are far from equilibrium processes.

\subsection{Treating the CTRW Equations}

\subsubsection{Remarks on the solution of the CTRW equations}

A standard way to treat the CTRW equation is by transforming them
to Fourier (F)
and Laplace (L) space,
whereby
the convolution theorems of the two transforms are used. We will
illustrate this procedure on the example of the waiting model
below in Sec.\ \ref{solvwait}.

The CTRW equations are though not always of a convolution type, e.g.\
the velocity model has not a convolution structure anymore, due to the
appearance of time in the integration limits (see Eq.\ (\ref{PVeq})), so that Fourier Laplace
methods are not directly applicable anymore, and other methods
are needed (actually also in the expression
for $Q$, Eq.\ (\ref{Qeq}), time appears in the $\D z$-integration limits in case of the velocity
model).

FL transforms, if applicable, usually do not allow to calculate the
probability $P(z,t)$ in closed
analytical form, but rather some asymptotic properties of it,
such as the mean square displacement at large times
\citep[e.g.][]{Klafter87,Blumen89}.
On the other hand, FL transforms, if applicable, allow to transform
the CTRW equations into other kinds of equations,
e.g.\  in one instead of the two
integral equations \citep[e.g.]{Klafter87,Blumen89},
into an integro-differential equation
\citep[or master equation, e.g.][]{Klafter87}, or even into
a fractional diffusion equation,
which has the form of a diffusion
equations, the fractional
derivatives are though generalized, non-local differentiation operators.
In Sec.\ \ref{fracdiffeq}, we will show how a fractional diffusion
equation arises naturally in the context of the waiting model
\citep[see also e.g.][]{Metzler00,Metzler04}.

Another standard way of treating the CTRW equations is with Monte Carlo simulations
\citep[see e.g.][]{Vlahos04}, or else,
the equations can be solved numerically, with an appropriate method
\citep{Isliker08}.

\subsubsection{Fourier and Laplace transforms of probability densities}

For any probability density function (pdf) such as $q_{\D z}$, we can define the Fourier transform
as ($z\to k$)
\beq
\hat{q}_{\D z}(k) = \int e^{-ik\D z}q_{\D z}(\D z) \, d\D z,
\eeq
which is often called the characteristic function of $q_{\D z}$.
Considering then the expression
\bea
i^n \p_k^n \hat{q}_{\D z}(k) \Big\vert_{k=0}
&=& i^n \int (-i\D z)^n e^{-ik\D z}q_{\D z}(\D z) \, d\D z \Big\vert_{k=0} \nonumber \\
&=& \int \D z^n q_{\D z}(\D z) \, d\D z  ,
\eea
for $n=0,1,2,3...$,
we see that, because $q_{\D z}$ is a pdf, the
last expression is the expectation value $\langle \D z^n \rangle$ of $\D z^n$,
the so-called $n$th moment, and
we have
\beq
i^n \p_k^n \hat{q}_{\D z} (0)
=\langle \D z^n \rangle   .
\label{dermo}
\eeq
In particular, it always holds that $\langle \D z^0 \rangle= 1$, since
$q_{\D z}$ is a pdf that is normalized to one
($\langle \D z^0 \rangle=\int q_{\D z}(\D z) \, d\D z=1$). Furthermore,
$\langle \D z^1 \rangle=\int \D z q_{\D z}(\D z) \, d\D z$
is the mean value of $q_{\D z}$.

In the use of Fourier and Laplace transforms for solving the CTRW
equations, we will concentrate on the asymptotic, large $|z|$ 
regime,
which corresponds to small values of $k$ (small
wave-numbers correspond to large length-scales or wave-lengths).
We thus can make a
Taylor expansion of $\hat{q}_{\D z}(k)$ around
$k=0$ and keep only a few low order terms,
\beq
\hat{q}_{\D z}(k) = \hat{q}_{\D z}(0) + \p_{k} \hat{q}_{\D z}(0)k
                 + \frac{1}{2} \p_{k}^2 \hat{q}_{\D z}(0) k^2 + ... ,
\eeq
With Eq.\ (\ref{dermo}),
the derivatives can be replaced with the moments,
\beq
\hat{q}_{\D z}(k) = 1 - i \langle \D z \rangle  k
                 - \frac{1}{2} \langle \D z^2 \rangle k^2 + ...
\label{TaylorF}
\eeq
(with $\langle \D z^0 \rangle = 1$). The
Taylor expansion of a pdf in terms of the moments is
practical because
the moments are the natural characteristics of a pdf.
Often, the distribution of spatial increments is assumed to be
symmetric around $z=0$, so that $\langle \D z \rangle = 0$, and the
Taylor expansion writes in this case as
\beq
\hat{q}_{\D z}(k) = 1
                 - \frac{1}{2} \langle \D z^2 \rangle k^2 + ...
\label{TaylorF2}
\eeq

Temporal distributions, such as the time-step distribution
$q_{\D t}(\D t)$ in the waiting model,
have the characteristic to be
'one-sided', i.e.\ they are defined and used only for $t\geq 0$, so that
it is more
appropriate to use Laplace transforms in this case, defined as
\beq
\tilde{q}_{\D t}(s) = \int_0^\infty e^{-s\D t} q_{\D t}(\D t)\, d\D t  ..
\eeq
Straightforward calculations and the use of the respective
definitions leads to the analogue of Eq.\ (\ref{dermo}) for
Laplace transforms,
\beq
(-\p_s)^n \tilde{q}_{\D t}(s) \Big\vert_{s=0}
=\int_0^\infty \D t^n q_{\D t}(\D t)\, d\D t = \langle \D t^n \rangle ,
\label{dermoL}
\eeq
where the $\langle \D t^n \rangle$ are again the moments.
As with respect to $z$, we will focus on the asymptotic, large $t$ regime,
which corresponds to small values of $s$,
and we make a Taylor-expansion of the Laplace transform around
$s=0$, replacing through Eq.\ (\ref{dermoL}) the derivatives by the
moments,
\beq
\tilde{q}_{\D t}(s) = 1 - \langle \D t \rangle s
+ ...  ,
\label{TaylorL}
\eeq
in complete analogy to Eq.\ (\ref{TaylorF}) ($\langle \D t^0 \rangle=1$ is
the normalization of $q_{\D t}$).
The Taylor expansions in Eqs.\ (\ref{TaylorF}), (\ref{TaylorF2}), and (\ref{TaylorL})
can of course only be used if the involved moments are finite.

\subsubsection{The symmetric and the one-sided Levy distributions\label{symmLevy}}

The {\it symmetric Levy distributions} are defined in Fourier space as
\beq
\hat{q}^{L,\alpha}_{\Delta z}(k) = \exp(-a\vert k\vert^\alpha) ,
\label{levydis}
\eeq
with $0<\alpha\leq 2$. It is not possible to express them in closed form
in real space, with two exceptions,
the case $\alpha=2$ is the usual Gaussian distribution
(the Fourier back-transform of a Gaussian is a Gaussian), and the
case $\alpha=1$ is known as the Cauchy distribution
\citep[see e.g.][Chap.\ 4.3]{Hughes95}. As mentioned
in Sec.\ \ref{subsub1}, the Levy distributions for $\alpha<2$
all have infinite
variance, and for $\alpha\leq 1$ they even have an infinite mean value,
so that the expansion in the form of Eq.\ (\ref{TaylorF2}) is not applicable.
We can though directly expand the exponential in Eq.\ (\ref{levydis}) and
find the small $k$ expansion as
\beq
\hat{q}^{L,\alpha}_{\Delta z}(k) = 1 -a\vert k\vert^\alpha
+ ... ,
\label{LexpaF}
\eeq
The case $\alpha=2$ corresponds obviously to the classical case
of Eq.\ (\ref{TaylorF2})
with finite second moment ($a\equiv (1/2)\langle \D z^2\rangle$) and zero
mean   (we are using only the symmetric Levy-distributions).

If one-sided distributions with infinite variance are needed, then they can also be defined
via Fourier space as a specific asymmetric Levy distribution,
or, more convenient for our purposes,
as shown in \citet{Hughes95} (Chap.\ 4.3.2), they can be defined
in Laplace space as
\beq
\tilde{q}^{L1,\beta}(s) = \exp(-b s^\beta) ,
\label{Llap}
\eeq
with $b$ strictly positive, and where now $0<\beta\leq 1$. These
one-sided Levy distributions decay as $t^{-1-\beta}$ for $t\to\infty$,
and they have the small $s$ expansion
\beq
\tilde{q}^{L1,\beta}(s) = 1- b s^\beta
+ ...
\label{LexpaL}
\eeq
Note that for $\beta=1$ we recover the finite mean case of
Eq.\ (\ref{TaylorL}), with $b=\langle \D t\rangle$,
and the back-transform of Eq.\ (\ref{Llap}) in this case
yields the distribution $\delta(\D t- b)$, i.e.\ the time-steps
are constant and equal to $b$ ($\Delta t\equiv \langle \D t\rangle\equiv b$).

In the following, we will use the (small $k$) Fourier expansion for $\hat{q}_{\D z}(k)$
in the form,
\beq
\hat{q}_{\D z}(k) = 1 -a\vert k\vert^\alpha + ... ,
\label{qexpaF}
\eeq
which for $\alpha<2$ corresponds to the Levy case, Eq.\ (\ref{LexpaF}),
and for $\alpha=2$ it recovers the normal, finite variance case of
Eq.\ (\ref{TaylorF2}), with $a=(1/2)\langle \D z^2\rangle$.
Correspondingly, the (small $s$) Laplace expansion for $\tilde{q}_{\D t}(s)$ will be used
in the form
\beq
\tilde{q}_{\D t}(s) = 1- b s^\beta  + ... ,
\label{qexpaL}
\eeq
which for $\beta<1$ yields the Levy distribution, Eq.\ (\ref{LexpaL}),
and for $\beta=1$ the normal, finite mean case of Eq.\ (\ref{TaylorL}),
with $b=\langle \D t\rangle$.

\subsubsection{Solving the CTRW equations with Fourier and Laplace transforms\label{solvwait}}

In this section, we will illustrate the use of the
Fourier and Laplace (F-L) transform to solve the CTRW
equation on the example of the waiting model, Eqs.\ (\ref{Qeq})
and (\ref{PWeq}).
The use of the respective convolution theorems, namely
\beq
\int f(x-y)g(y) \,dy \to \hat{f}(k)\,\hat{g}(k)
\eeq
for Fourier transforms, and
\beq
\int_0^t \phi(t-\tau)\psi(\tau) \,d\tau \to \tilde{\phi}(s)\,\tilde{\psi}(s)
\eeq
for Laplace transforms (with $f,\,g,\,\phi$, and $\psi$ any
transformable functions),
allows in the
case of the waiting model to determine
the solution in Fourier Laplace space: For the initial condition
$P(z,t=0)=\delta(t)\delta(z)$ and in the absence of any source ($S=0$),
Eq.\ (\ref{Qeq}) turns into
$\tilde{\hat{Q}}(k,s)=\tilde{\hat{Q}}(k,s)\,\hat{q}_{\D z}(k)\,\tilde{q}_{\D t}(s)+1$, and
Eq.\ (\ref{PWeq}) takes the form
$\tilde{\hat{P}}_W(k,s)=\tilde{\hat{Q}}(k,s)\tilde{\Phi}_W(s)$,
and we can eliminate $\tilde{\hat{Q}}$ and solve for $\tilde{\hat{P}}$. Noting
further that $\Phi_W(t)=\int_t^\infty q_{\D t}(\D t)\,d\D t
=1-\int^t_\infty q_{\D t}(\D t)\,d\D t$, so that
$\tilde{\Phi}_W(s)=(1-\tilde{q}_{\D t}(s))/s$, we find
\beq
\tilde{\hat{P}}_W(k,s) = \frac{1-\tilde{q}_{\D t}(s)}{s\left[1-\hat{q}_{\D z}(k)
\tilde{q}_{\D t}(s)\right]}  ,
\label{mowei}
\eeq
which is known as the {\it Montroll-Weiss equation} \citep[e.g.\ ][]{Montroll65,Zumofen93,Klafter87}.

Looking for asymptotic solutions, we insert the general form
of the transformed temporal and spatial step distribution,
Eqs.\ (\ref{qexpaF}) and (\ref{qexpaL}), respectively,
into Eq.\ (\ref{mowei}), which yields
\beq
\tilde{\hat{P}}_W(k,s) = \frac{b s^{\beta-1}}
{b s^\beta +a\vert k\vert^\alpha}  .
\label{PWsmall}
\eeq
Unfortunately, it is not possible to
Fourier and Laplace back-transform $\tilde{\hat{P}}_W(k,s)$ analytically.
We can though use Eq.\ (\ref{dermo}) for $n=2$, namely
\beq
\langle z^2(s) \rangle
=
- \p_k^2 \tilde{\hat{P}}_W(k=0,s) ,
\label{msdfou}
\eeq
to determine the mean square
displacement in the asymptotic
regime
(note that we set $k=0$ at the end, which clearly is in the large $|z|$
regime).
Inserting $\tilde{\hat{P}}_W(k,s)$ from Eq.\ (\ref{PWsmall})
into Eq.\ (\ref{msdfou}),
without yet setting $k=0$, we find
\beq
\langle z^2(s)\rangle = -\frac{2a^2\alpha^2 \vert k\vert^{2\alpha-2}}{bs^{2\beta+1}}
+\frac{a \alpha(\alpha-1) \vert k\vert^{\alpha-2}}{bs^{\beta+1}}
\label{aux1}
\eeq
The first term on the right side diverges for $\alpha<1$, and the second term
diverges for $\alpha<2$, so that $\langle z^2(s)\rangle$ is infinite in these
cases. This divergence must be interpreted in the sense that the diffusion
process is very efficient, so that in the asymptotic
regime, $P_W$ has
already developed  so fat wings at large $|z|$ (power-law tails)
that the variance, and with that the mean square displacement,
of $P_W$ is infinite, $P_W$ has already become a Levy type
distribution \citep[see also][]{Klafter87,Balescu07}.
Of course, with the formalism we apply we cannot say anything
about the transient phase, before the asymptotic regime
is reached.
We just note here that the velocity model (Eq.\ (\ref{PVeq}))in this regard
is not so over-efficient, it allows
super-diffusion with a more gradual build-up of the fat wings of the
distribution $P_V$ \citep{Klafter87}.

Less efficient diffusion can only be achieved
in the frame of the waiting model for $\alpha=2$, i.e.\ for
normal, Gaussian distributed spatial steps (see Eq.\ (\ref{qexpaF})). In this case,
Eq.\ (\ref{aux1}) takes the form
\beq
\langle z^2(s)\rangle = \frac{\langle\D z^2\rangle}{bs^{\beta+1}}
\eeq
($a=(1/2)\langle\D z^2\rangle$ for $\alpha=2$).
This expression is valid for small $s$,
and with the help of the
Tauberian theorems, which relate the power-law scaling
of a Laplace transform at small $s$ to the scaling in original
space for large $t$ \citep[see e.g.][]{Hughes95,Feller71} it follows that
\beq
\langle z^2(s)\rangle\rangle \sim t^\beta .
\eeq
With our restriction $0<\beta\leq 1$, diffusion is always
of sub-diffusive character, and for $\beta=1$ it is normal,
as expected, since we have in this case waiting times with finite
mean and variance (see
Eq.\ (\ref{qexpaL})).


\subsection{Including Velocity Space Dynamics\label{velo}}

Above all in applications to turbulent systems, and mainly to turbulent or driven
plasma systems, it may not be enough to monitor the position and the timing of
a particle, since its velocity may drastically change, e.g.\ if it interacts with
a local electric field generated by turbulence.
An interesting extension of the standard CTRW for these cases is to include, besides the
position space and temporal dynamics, also the velocity space dynamics, which
allows to study anomalous diffusive behaviour also in energy space.

To formally define the extended CTRW that also includes momentum space,
we keep Eq.\ (\ref{rnclass}) and Eq.\ (\ref{tctrw}) for position and
time evolution as they
are, and newly the momentum (or velocity) also becomes a random,
dynamic variable,
with temporal evolution of the form
\beq
\vec p_n = \Delta \vec p_n + \Delta \vec p_{n-1} + \Delta \vec p_{n-2} + \, ...
                                             \, + \Delta \vec p_1 + \vec p_0 ,
\label{pext}
\eeq
with $p_0$ the initial momentum, and the $\D \vec p_i$ the momentum increments.
Again,
one has to specify a functional form for the distribution of momentum
increments
$q_{\D \vec p}(\D \vec p)$ in order to specify the random walk problem
completely. The solution of the extended CTRW is in the form of
the distribution $P(\vec r,\vec p,t)$ for a particle at time $t$ to be
at position $\vec r$ and to have momentum $\vec p$.

The extended CTRW can be treated by Monte-Carlo simulations,
as done in \citet{Vlahos04},
or in \citet{Isliker08} a set of equations for the extended CTRW
has been introduced, which basically is a generalization of
Eq.\ (\ref{Qeq}) and Eq.\ (\ref{PVeq}), and a way to solve
the equations numerically is presented.

\section{From random walk to fractional diffusion equations}

The purpose of this section is to show how fractional diffusion equations
naturally arise in the context of random walk models.
The starting point here are the CTRW equations for the waiting
model, Eqs.\ (\ref{Qeq}) and (\ref{PWeq}), which in Fourier Laplace
space take the form of Eq.\ (\ref{mowei}), and on inserting
the small $k$ and small $s$ expansion of the step-size and waiting-time
distributions, Eqs.\ (\ref{qexpaF}) and (\ref{qexpaL}), respectively,
the waiting CTRW equation
takes the form of Eq.\ (\ref{PWsmall}), with $\alpha\leq2$ and $\beta \leq 1$.
Multiplying Eq.\ (\ref{PWsmall}) by the numerator on its right side,
\beq
\tilde{\hat{P}}_W(k,s)(b s^\beta +a\vert k\vert^\alpha) = b s^{\beta-1}  ,
\label{PWsmall2}
\eeq
and rearranging, we can bring the equation for $\tilde{\hat{P}}_W$ to the form
\beq
s^\beta \tilde{\hat{P}}_W(k,s) - s^{\beta-1}
= -\frac{a}{b} \vert k\vert^\alpha \tilde{\hat{P}}_W(k,s)  .
\label{PWsmall3}
\eeq

It is illustrative to first consider the case of normal diffusion,
with $\beta=1$ and $\alpha=2$,
where according
to Eqs.\  (\ref{qexpaF}) and (\ref{qexpaL}) we have $a=(1/2)\langle \D z^2\rangle$ and
$b=\langle \D t\rangle$, so that
\beq
s \tilde{\hat{P}}_W(k,s) - s^{0}  = -\frac{\langle \D z^2\rangle}{2\langle \D t\rangle}
\vert k\vert^2 \tilde{\hat{P}}_W(k,s)
\label{PWsmall4}
\eeq
Now recall how a first order temporal derivative is expressed in Laplace space,
\beq
\frac{d}{dt} \psi(z) \to s \tilde{\psi}(s) - s^{0}\psi(0) ,
\label{derivL}
\eeq
and a how a spatial derivative translates to Fourier space,
\beq
\frac{d^n}{dz^n} f(z) \to (-ik)^n \hat{f}(k)
\label{derivF}
\eeq
Obviously, for $P_W(z,t=0)=\delta(z)$, Eq.\ (\ref{PWsmall4}) can be
back-transformed as
\beq
\p_t P_w(z,t) = \frac{\langle \D z^2\rangle}{2\langle \D t\rangle}
                 \p^2_z P_W(z,t),
\eeq
so that we just recover the simple diffusion equation of the normal
diffusive case.

\subsection{Fractional derivatives}

Fractional derivatives are a generalization of the usual derivatives
of $n$th order to general non-integer orders. There exist several definitions,
and in original space ($z$ or $t$) they are a combination of usual derivatives
of integer order and integrals over space (or time). The latter property makes
them non-local operators, so that fractional differential
equations are non-local equations, as are the CTRW integral equations.
For the following, we need to define the Riemann-Liouville left-fractional
derivative of order $\alpha$,
\beq
{}_a D^\alpha_{z} f(z)
=
\frac{1}{\Gamma(n-\alpha)}
\frac{d^n}{dz^n} \int_a^z \frac{f(z^\prime)}{(z-z^\prime)^{\alpha+1-n}}
\,dz^\prime  ,
\eeq
with $\Gamma$ the usual Gamma-function, $a$ a constant, $n$ an integer
such that $n-1\leq\alpha<n$, $\alpha$ a positive real number,
and $f$ any suitable function.
Correspondingly, the Riemann-Liouville right-fractional
derivative of order $\alpha$ is defined as
\beq
{}_z D^\alpha_{b} f(z)
=
\frac{(-1)^n}{\Gamma(n-\alpha)}
\frac{d^n}{dz^n} \int_z^b \frac{f(z^\prime)}{(z^\prime-z)^{\alpha+1-n}} \,dz^\prime .
\eeq
It is useful to combine these two asymmetric definitions into a
a new, symmetric fractional derivative, the so-called
Riesz fractional derivative,
\beq
D^\alpha_{|z|} f(z)=
-\frac{1}{2\cos(\pi\alpha/2)}
\left({}_{-\infty} D^\alpha_{z}  + {}_z D^\alpha_{\infty} \right) f(z)   ..
\eeq
The Riesz fractional derivative has the interesting property that its representation in Fourier
space is
\beq
{}^{(R)}D^\alpha_{\vert z \vert} f(z) \to -\vert k\vert^\alpha \hat{f}(k) .
\label{RieszF}
\eeq
Comparison of this simple expression with Eq.\ (\ref{derivF})
makes obvious that the Riesz derivative is a natural generalization
of the usual derivative with now non-integer $\alpha$.

To treat time, a different variant of fractional derivative
is useful, the
Caputo fractional derivative of order $\beta$,
\beq
{}^{(C)}D_t^\beta \psi(t)
=
\frac{1}{\Gamma(n-\beta)}
 \int_0^t \frac{1}{(t-t^\prime)^{\beta+1-n}} \frac{d^n}{dt^{\prime n}} \psi(t^\prime)\,dt^\prime ,
\eeq
with $n$ an integer such that $n-1\leq\beta<n$, and
$\psi$ any appropriate function.
The Caputo derivative
translates to Laplace space as
\beq
{}^{(C)}D_t^\beta \psi(t) \to s^\beta \tilde{\psi}(s) - s^{\beta-1}\psi(0) ,
\label{CapF}
\eeq
for $0<\beta\leq 1$,
which is again a natural generalization of Eq.\ (\ref{derivL}) for
the usual derivatives for now non-integer $\beta$ (the Caputo derivative
is also defined for $\beta\geq1$, with Eq.\ (\ref{CapF}) taking a more
general form).

Further details about fractional derivatives can be found e.g.\
in \citet{Podlubny99} or in the extended Appendix of \citet{Balescu07b}.

\subsection{Fractional diffusion equation\label{fracdiffeq}}

Turning now back to Eq.\ (\ref{PWsmall3}), we obviously can identify the
fractional Riesz and Caputo derivatives in their simple Fourier and
Laplace transformed form, Eqs. (\ref{RieszF}) and (\ref{CapF}), respectively,
and write
\beq
{}^{(C)}D_t^\beta P_W(z,t) = \frac{a}{b}\,\,\,
                  {}^{(R)}D^\alpha_{\vert z \vert} P_W(z,t)
\label{fraceq}
\eeq
>From this derivation it is clear that the order of the fractional derivatives,
$\alpha$ and $\beta$, are determined by the index of the
step-size ($q_{\D z}$) and the waiting time ($q_{\D t}$)
Levy distributions, respectively.
It is also clear that Eq. (\ref{fraceq}) is just an alternative way of writing
Eq.\ (\ref{PWsmall3}) or (\ref{PWsmall}), and as such it is the asymptotic,
large $|z|$, large $t$ version of the CTRW equations
(\ref{Qeq}) and (\ref{PWeq}).
It allows though to apply different mathematical tools for its
analysis that have been
developed specially for fractional differential equations.


As an example, we may consider the case $\beta=1$ and $0<\alpha\leq 2$,
where the diffusion equation is fractional just in the spatial part,
\beq
\p_t P_w(z,t) = \frac{a}{b}\,\,\,
                  {}^{(R)}D^\alpha_{\vert z \vert} P_W(z,t)
\label{fraceqexam}
\eeq
In Fourier Laplace space, this equation takes the form
\beq
\hat{P}_W(k,s) = \frac{b}
{b s +a\vert k\vert^\alpha}  ,
\label{PWexam}
\eeq
which, on applying the inverse Laplace transform, yields
\beq
\hat{P}_W(k,t) = \exp\left(- \frac{a}{b}\, |k|^\alpha t\right) ,
\eeq
which is the Fourier-transform of a symmetric Levy-distribution
with time as a parameter (see Eq.\ (\ref{levydis})),
and with index $\alpha$ equal to the
one of the spatial step distribution $q_{\D z}$.
Thus, for $\alpha<2$, the
solution has power-law tails, and the mean square displacement
(or variance or second moment)
is infinite, as we had found it in Eq.\ (\ref{aux1}).
For $\alpha=2$, the solution $P_W(z,t)$ is a Gaussian
(the Fourier back-transform of a Gaussian is a Gaussian),
and we have normal diffusion.

\section{Action diffusion in Hamiltonian systems\label{hamdiff}}

So-far, our starting point for modeling diffusion was mostly the random walk approach
and a probabilistic equation of the Chapman Kolmogorov type (Eq.\ (\ref{chapkol})). Here now,
we turn to Hamiltonian systems, and we will show how from Hamilton's equations a quasi-linear
diffusion equation can be derived. This diffusion equation is of practical interest when
the Hamiltonian system consists in a large number of particles, so that it becomes technically
difficult to follow the individual evolution of all the particles.

Let us consider a generic $N-$degrees of freedom Hamiltonian system with Hamiltonian $H\left(\bold{q},\bold{p}\right)$ and equations of motion given by
\begin{eqnarray}
\frac{d\bold{q}}{dt}&=&\frac{\partial H}{\partial \bold{p}}, \\
\frac{d\bold{p}}{dt}&=&-\frac{\partial H}{\partial \bold{q}},
\end{eqnarray}
where $\bold{q}=\left(q_1,...,q_N \right)$ and $\bold{p}=\left(p_1,...,p_N \right)$ are the canonical coordinates and momenta, respectively. In order to be integrable such a system should have $N$ independent invariants of the motion, corresponding to an equal number of symmetries of the system \citep{Goldstein}. The integrability of a system is a very strong condition which does not hold for most systems of physical interest. However, in most cases we can consider our system as a perturbation of an integrable one and split the Hamiltonian accordingly to an integrable part and a perturbation. Then the description of the system can be given in terms of the action-angle variables of the integrable part (Note that a periodic integrable system can always be transformed to action-angle variables \citep{Goldstein}), so that
we can write
\begin{equation}
H(\bold{J},\mbox{\boldmath$\theta$},t)=H_0(\bold{J})+\epsilon H_1(\bold{J},\mbox{\boldmath$\theta$},t)  ,
\end{equation}
with $\bold{J}=\left(J_1,...,J_N\right)$ and $\mbox{\boldmath$\theta$}=\left(\theta_1,...\theta_N\right)$ being the action and angle variables, respectively. $H_0$ is the integrable part of the original Hamiltonian and $H_1$ is the perturbation.  The parameter $\epsilon$ is dimensionless and will be used only for bookkeeping purposes in the perturbation theory; it can be set equal to unity, in the final results. The evolution of the integrable system $H_0$
is given by the following equations of motion
\begin{eqnarray}
\dot{\bold{J}}&=&0 \\
\dot{\mbox{\boldmath$\theta$}}&=&\mbox{\boldmath$\omega_0$}t+\mbox{\boldmath$\theta_0$} ,
\end{eqnarray}
where $\mbox{\boldmath$\omega_0$}=\partial H_0/\partial\bold{J}$ are the frequencies of the integrable system $H_0$. The N action variables correspond to the $N$ invariants of the motion required for the integrability of the system. \

The perturbation $H_1$ leads to the breaking of this invariance due to its $\mbox{\boldmath$\theta$}$ dependence. The derivation of a quasilinear diffusion equation in the action space is the subject of this section, and the method to be used is the canonical perturbation theory applied for finite time intervals. This method of derivation is based on first principles and does not imply any statistical assumptions for the dynamics of the system, such as the presence of strong chaos resulting in phase mixing or loss of memory for the system. Moreover, it is as systematic as the underlying perturbation scheme of the canonical perturbation theory, so it can be extended to higher order and provide results beyond the quasilinear approximation \citep{PRE}. Also, it is important to note that the method makes quite clear what physical effects are taken into account in the quasilinear limit and what effects are actually omitted. It is worth mentioning that non-quasilinear diffusion has been studied both analytically and numerically for a variety of physical systems \citep{CaEsVe_90,HeKj_94,BeEs_98,LaPe_83b,LaPe_99}.\

The basic idea of the canonical perturbation theory is the search of canonical transformations for the perturbed system (i.e. transformations which preserve the Hamiltonian structure of the system) under which the new (transformed) Hamiltonian is a function of the action only. For a near-integrable system this can be done approximately, and the new actions correspond to approximate invariants of the motion which contain all the essential features of the phase space structure. The  transformations involved in canonical perturbation theory are expressed in terms of the so-called mixed-variable generating functions. These can be functions of a subset of the old variables along with a subset of the new ones \citep{Goldstein}. Thus, the transformation from $(\bold{J},\mbox{\boldmath$\theta$})$ to $(\bold{\bar{J}}, \mbox{\boldmath$\bar{\theta}$})$ can be expressed by a generating function of the form $S(\bold{\bar{J}}, \mbox{\boldmath$\theta$},t)$. The transformation equations are given in the following implicit form:
\begin{eqnarray}
\bold{J}&=&\bold{\bar{J}}+\epsilon \frac{\partial S(\bold{\bar{J}},\mbox{\boldmath$\theta$},t)}{\partial \mbox{\boldmath$\theta$}} \label{JJ_0}  ,\\
\mbox{\boldmath$\bar{\theta}$}&=&\mbox{\boldmath$\theta$}+\epsilon \frac{\partial S(\bold{\bar{J}},\mbox{\boldmath$\theta$},t)}{\partial \bold{\bar{J}}}\label{thetatheta_0} .
\end{eqnarray}

Following a standard procedure \citep{Goldstein}, we seek a transformation to new variables $(\bar{J},\bar{\theta})$ for which the new Hamiltonian $\bar{H}$ is a function of the action $\bold{\bar{J}}$ alone. Expanding $S$ and $\bar{H}$ in power series of a small parameter $\epsilon$
\begin{eqnarray}
S&=&\bold{\bar{J}}\mbox{\boldmath$\theta$}+\epsilon S_1+\epsilon^2 S_2+...\\
\bar{H}&=&\bar{H}_0+\epsilon \bar{H}_1+\epsilon^2 \bar{H}_2+...  ,
\end{eqnarray}
where the lowest-order term has been chosen to generate the identity transformation $\bold{J}=\bold{\bar{J}}$ and $\mbox{\boldmath$\bar{\theta}$}=\mbox{\boldmath$\theta$}$. The old action and angle can be also expressed as power series in $\epsilon$:
\begin{eqnarray}
\bold{J}&=&\bold{\bar{J}}+\epsilon \frac{\partial S_1(\bold{\bar{J}},\mbox{\boldmath$\theta$},t)}{\partial \mbox{\boldmath$\theta$}}+\epsilon^2 \frac{\partial S_2(\bold{\bar{J}},\mbox{\boldmath$\theta$},t)}{\partial \mbox{\boldmath$\theta$}}+ ...  \label{JJ}\\
\mbox{\boldmath$\bar{\theta}$}&=&\mbox{\boldmath$\theta$}+\epsilon \frac{\partial S_1(\bold{\bar{J}},\mbox{\boldmath$\theta$},t)}{\partial \bold{\bar{J}}}+\epsilon^2 \frac{\partial S_2(\bold{\bar{J}},\mbox{\boldmath$\theta$},t)}{\partial \bold{\bar{J}}}+ ...  ,\label{thetatheta}
\end{eqnarray}
and the new Hamiltonian is
\begin{equation}
\bar{H}(\bold{\bar{J}},\mbox{\boldmath$\bar{\theta}$},t)=H(\bold{J},\mbox{\boldmath$\theta$},t)+\frac{\partial S(\bold{\bar{J}},\mbox{\boldmath$\theta$},t)}{\partial t} . \label{Hbar}
\end{equation}
By substituting the respective power series in Eq.\ (\ref{Hbar}) and equating like powers of $\epsilon$ for the zero order we have
\begin{equation}
\bar{H}_0=H_0  ,
\end{equation}
while in the first and second order we have the equations
\begin{equation}
\frac{\partial S_i}{\partial t}+\mbox{\boldmath$\omega_0$}\frac{\partial S_i}{\partial \mbox{\boldmath$\theta$}}=\bar{H}_i-F_i(\bold{J},\mbox{\boldmath$\theta$},t),  \hspace{3em} i=1,2 ,
\end{equation}
with
\begin{eqnarray}
F_1(\bold{J},\mbox{\boldmath$\theta$},t)&=&H_1 ,\\
F_2(\bold{J},\mbox{\boldmath$\theta$},t)&=&\frac{1}{2}\frac{\partial^2 H_0}{\partial\bold{J}^2}\left(\frac{\partial S_1}{\partial\mbox{\boldmath$\theta$}}\right)^2 +\frac{\partial H_1}{\partial \bold{J}}\frac{\partial S_1}{\partial \mbox{\boldmath$\theta$}}
\end{eqnarray}
providing the first and second order generating function $S_1$ and $S_2$, respectively. The latter are  linear partial equations which can be solved in a time interval of interest $[t_0,t]$ by the method of characteristics (i.e. integration along the unperturbed orbits). Note that $\bar{H}_1$ and $\bar{H}_2$ are arbitrary functions which can be set equal to zero (for the application of the canonical perturbation theory in infinite time intervals, these functions have to be chosen so that they cancel secular terms \citep{Goldstein}).
For a general perturbation of the form
\begin{equation}
H_1=\sum_{\bold{m}\neq 0}H_\bold{m}(\bold{J},t)e^{i\bold{m}\cdot\mbox{\boldmath$\theta$}} ,
\end{equation}
the solution for the first order generating function $S_1$ can be written as
\begin{equation}
S_1(\bold{\bar{J}}, \mbox{\boldmath$\theta$},t;t_0)=-\sum_{\bold{m}\neq 0}e^{i\bold{m}\cdot\left(\mbox{\boldmath$\theta$}-\mbox{\boldmath$\omega_0$}t\right)}\int_{t_0}^tH_\bold{m}(\bold{J},s)e^{i\bold{m}\cdot\mbox{\boldmath$\omega_0$}s}ds . \label{S1}
\end{equation}
Similarly, the solution for $S_2$ can be readily obtained. The resulting expression (too lengthy to be presented here) is a periodic function of $\mbox{\boldmath$\theta$}$. This is the only information we need for $S_2$, since its exact form will not be involved in our calculations.\

In the following, we show that the results of first order perturbation theory can be utilized in order to provide an evolution equation for the angle-averaged distribution function, which is accurate up to second order with respect to the perturbation parameter $\epsilon$, namely a quasilinear action diffusion equation. Therefore, we can relate results from perturbation theory applied for a single particle motion, to the distribution function, describing collective particle motion. The latter is of physical interest in all cases where a large number of particles is involved in collective phenomena so that a statistical approach is required.\

The evolution of the phase space distribution function $F$ is governed by Liouville's equation \citep{Goldstein}
\begin{equation}
\frac{\partial F}{\partial t}+[F,H]=0 ,
\end{equation}
where $[.,.]$ denotes the Poisson bracket, defined as $[f_1,f_2]=\nabla_\bold{q}f_1\cdot\nabla_\bold{p}f_2-\nabla_\bold{p}f_1\cdot\nabla_\bold{q}f_2$ with $\bold{q}$ and $\bold{p}$ being the canonical positions and momenta, respectively. This equation simply expresses the incompressibility of the Hamiltonian flow and the invariance of the number of particles. It is well-known that for an integrable system, any function of the invariants of the motion (actions) is a solution of the Liouville's equation. For the case of a near-integrable system, an approximate distribution function can be obtained as a function of the approximate invariant of the motion, namely the new actions $\bold{\bar{J}}$, so that we can write
\begin{equation}
F(\bold{J},\mbox{\boldmath$\theta$},t)=F(\bold{\bar{J}}) ,\label{F_Jbar}
\end{equation}
with $\bold{\bar{J}}$ given implicitly by Eq.\ (\ref{JJ}). To second order, with respect to $\epsilon$ we have
\begin{equation}
\bold{\bar{J}}=\bold{J}-\epsilon\Delta_1 \bold{J}+\frac{\epsilon^2}{2}\frac{\partial}{\partial \bold{J}}\cdot\left(\Delta_1 \bold{J}\Delta_1 \bold{J}\right) +\epsilon^2 \Delta_2 \bold{J} \label{Jbar} ,
\end{equation}
with
\begin{equation}
\Delta_i \bold{J}(\bold{J}, \mbox{\boldmath$\theta$},t;t_0)=\frac{\partial S_i(\bold{J}, \mbox{\boldmath$\theta$},t;t_0)}{\partial \mbox{\boldmath$\theta$}}, \hspace{3em} i=1,2. \label{D1J}
\end{equation}
Substituting (\ref{Jbar}) in (\ref{F_Jbar}) and utilizing a Taylor expansion with respect to $\epsilon$ we have
\begin{eqnarray}
F(\bold{J},\mbox{\boldmath$\theta$},t)
&=&
F(\bold{J}) - \epsilon\frac{\partial F(\bold{J})}{\partial \bold{J}} \cdot\left(\Delta_1 \bold{J}\right)
+ \frac{\epsilon^2}{2}\frac{\partial}{\partial \bold{J}}\cdot\left[\left(\Delta_1 \bold{J}\Delta_1 \bold{J}\right)\cdot\frac{\partial F(\bold{J})}{\partial \bold{J}}\right] \nonumber \\
&& + \epsilon^2 \frac{\partial F(\bold{J})}{\partial \bold{J}}(\Delta_2 \bold{J}) .
\end{eqnarray}
Noting that $\Delta_i \bold{J}(\bold{J}, \mbox{\boldmath$\theta$},t_0;t_0)=0$ we have $F(\bold{J},\mbox{\boldmath$\theta$},t_0)=F(\bold{J})$, and by averaging over the angles, we obtain
\begin{equation}
f(\bold{J},t)=f(\bold{J},t_0)+\frac{\epsilon^2}{2}\frac{\partial}{\partial \bold{J}}\cdot\left[\left<\left(\Delta_1 \bold{J}\Delta_1 \bold{J}\right)\right>_{\mbox{\boldmath$\theta$}}\cdot\frac{\partial f(\bold{J},t_0)}{\partial \bold{J}}\right] ,\label{f1}
\end{equation}
where $f$ is the angle-averaged distribution function
\begin{equation}
f(\bold{J},t)=\left<F(\bold{J},\mbox{\boldmath$\theta$},t)\right>_{\mbox{\boldmath$\theta$}} ,
\end{equation}
and we have used the fact that $\left<\Delta_i\bold{J}\right>_ {\mbox{\boldmath$\theta$}}=0$ as obtained from Eqs.\ (\ref{D1J}) and the fact that $S_i$ are sinusoidal functions of $\mbox{\boldmath$\theta$}$ . Taking the limit $t\rightarrow t_0$ in Eq.\ (\ref{f1}), we finally obtain the quasilinear action diffusion equation
\begin{equation}
\frac{\partial f}{\partial t}=\epsilon^2\frac{\partial}{\partial \bold{J}}\cdot\left[\bold{D}(\bold{J},t)\cdot\frac{\partial f}{\partial \bold{J}}\right] , \label{f2}
\end{equation}
with
\begin{equation}
\bold{D}(\bold{J},t)=\frac{1}{2}\frac{\partial\left<\left(\Delta_1 \bold{J}\Delta_1 \bold{J}\right)\right>_{\mbox{\boldmath$\theta$}}}{\partial t} \label{D}
\end{equation}
being the corresponding quasilinear diffusion tensor. Identifying that $\Delta_1 \bold{J}$ corresponds to the first order action variation, we see that $\bold{D}$ in Eq.\ (\ref{D}) corresponds to the common definition of the diffusion tensor, as provided by the statistical approach, which is based on the Kramers-Moyal expansion of the master equation \citep[see Sec.\ \ref{fokkpla}, and][]{VanKampen}.

It is worth mentioning that this form [Eq.\ (\ref{f2})] of the action diffusion 
equation is similar to the diffusion equation obtained with the utilization 
of the Fick's Law [Eq.\ (\ref{diffusioneq})]. However, it has been shown that for any 
Hamiltonian system, this form of the action diffusion equation is equivalent 
to the Fokker-Planck equation [Eq.\ (\ref{FokkerP})], due to the fact that the 
corresponding drift velocity and diffusion terms are related through
\beq
\bold{V} = \frac{\bold{D}}{\bold{z}} ,
\eeq
where $\bold{z}=\bold{J}$ for the case of the action diffusion equation 
\citep[see Chap.\ 4 in ][and references therein]{Lichtenberg92} . This property is 
implied directly from the canonical form of the underlying equations of  
motion. More generally, it has been shown that this form of the diffusion 
equation has further relation with the more general class of the 
microscopically reversible systems \citep[see][and references therein]{Molvig84}.

It is important to emphasize that the derivation procedure described here does not prerequisite any statistical assumption and is only based on the underlying equations of motion. In contrast to other statistical approaches, it is not necessary to assume strong stochasticity related to the completely chaotic regime where loss of memory takes place and the orbits are completely decorrelated. Therefore, Eq.\ (\ref{f2}), is capable of describing not only diffusion in the almost homogeneous "chaotic sea" in the phase space, but also intermittent motion in an inhomogeneous phase space structure where resonant islands and chaotic areas are interlaced. It is worth mentioning that the diffusion tensor $\bold{D}$ is time-dependent. The dependence of $\bold{D}$ on the actions, for the case of time-periodic perturbations, is through smooth localized functions which are centered around the corresponding resonances \citep{PRE}. In the limit $t\rightarrow \infty$ these localized functions tend to Dirac delta functions, which commonly appear in standard derivations of the quasilinear diffusion tensors \citep{Zaslavski}. However, the consideration of this limit in the derivation of $\bold{D}$ corresponds to an extension to infinity of the limits of integration in the derivation of $S_1$, in Eq.\ (\ref{S1}), which implies an ergodicity assumption and a steady-state approach. In the most general case, the diffusion tensor $\bold{D}$ [Eq.\ (\ref{D})] is capable of describing not only steady-state but also transient diffusion phenomena, through its time dependence. In this case the time scales of the action diffusion are determined by the time-dependence of $\bold{D}$ as well as by the factor $\epsilon^2$ in the r.h.s. of Eq.\ (\ref{f2}), which implies an actually slow diffusion process. Note that these time scales come naturally into play and there is no need for a priori separation of the distribution function in slow and fast varying parts as in many heuristic derivations of the Fokker-Planck equation.\

In the previous paragraphs we have derived a general quasilinear action diffusion equation for a Hamiltonian system, with a minimum of assumptions on the underlying dynamics. The fact that this method of derivation is closely related to a systematic and rigorous perturbation scheme allows for extending these results beyond the quasilinear approximation. Therefore we can carry our perturbation scheme to higher order in order to provide a hierarchy of diffusion equations having higher-order derivatives of the action distribution function with respect to the action, in direct analogy to statistical approaches where higher-order Kramers-Moyal expansions are considered \citep[see Sec.\ \ref{fokkpla}, and][]{VanKampen}. Note that the most appropriate method for handling calculations involved in higher-order perturbation theory is the utilization of Lie transforms \citep{PRE}. The hierarchy of higher-order diffusion equations does not only provide better accuracy with respect to the perturbation parameter $\epsilon$, but is also capable of describing non-Gaussian evolution of the distribution function and resonant processes between the particle and beats of multiple spectral components of the perturbation, known as nonlinear resonances.

\section{Other ways to model anomalous diffusion}

\citet{Escande07} review and discuss under what
circumstances the Fokker-Planck equation (Eq.\ (\ref{FokkerP}))
is able to model anomalous diffusion.
In summary, the FP equation, which is a local model, is
able to model anomalous diffusive behaviour
in cases where there is a non-zero drift velocity, $V(z)\ne 0$,
anomalous diffusion is thus based on drift effects.

\citet{Klafter87} briefly review attempts of using the Fokker-Planck
equation with zero drift velocity, but spatially or temporally dependent
diffusion coefficient. In order to account for anomalous
diffusion, the spatial or temporal dependence of the diffusion coefficient
must though be chosen in very particular ways, which are difficult
to interpret physically.

\citet{Lenzi03} shortly discuss
non-linear diffusion equations, which have the form of the simple
diffusion equation as in Eq.\ (\ref{einsdiff}), with $P$ though raised
on one side of the equation to some power $\gamma$.

\section{Applications in Physics and Astrophysics}

CTRW has successfully been applied to model various phenomena of anomalous
diffusion, including sub- and super-diffusive phenomena, in the
fields of physics, chemistry, astronomy, biology, and economics
\citep[see the references in][]{Metzler00,Metzler04}.

Laboratory plasma in fusion devices (tokamaks) show a variety of anomalous diffusion phenomena.
\citet{Balescu95} was the first to apply CTRW
to plasma physical problems. Later, \citet{vanMil04} and \cite{vanMil04b} developed
a CTRW model for confined plasma, the critical gradient model,
which was able to explain observed anomalous diffusion phenomena such as 'up-hill' transport,
where particles diffuse against the driving gradient.
\citet{Isliker08} studied the same physical system, with the use though of the extended
CTRW that includes momentum space dynamics, and they studied the evolution of the
density and temperature distribution and the particle and heat diffusivities.

Also in astrophysical plasmas anomalous diffusion is ubiquitous, there are many
astrophysical systems where non-thermal (i.e.\ not Maxwellian distributed) particles
are directly or indirectly, through their emission, observed.

\begin{figure}[ht]
\centering
  \includegraphics[width=5cm]{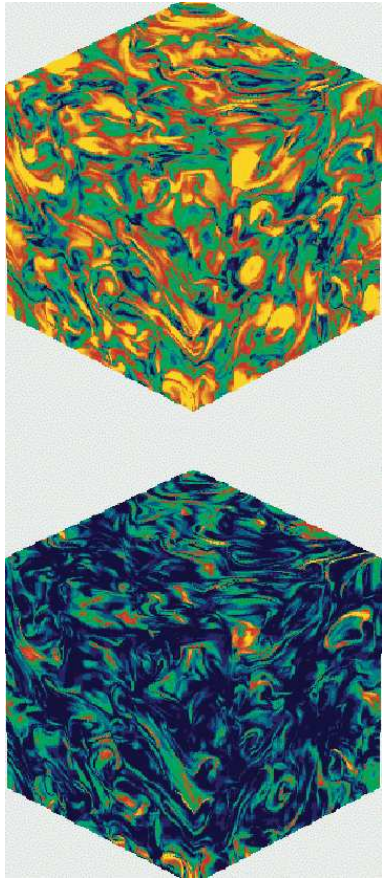}\\
  \caption{Visualization of the turbulent magnetic field $\mid B \mid$ (top) and electric field $\mid E \mid$ (bottom) in the simulation box. High values are in yellow (light) and low values in blue (dark).  }\label{turb}
\end{figure}

\cite{Dmitr1,Dmitr2} analyzed the acceleration of particles inside 3-D MHD turbulence. The compressible MHD equations  were solved numerically. In these simulations, the decay of large amplitude waves was studied. After a very short time (a few Alfv\'en times), a fully turbulent state with a broad range of scales has been developed (Fig. \ref{turb}).

\begin{figure}[ht]
\centering
  \includegraphics[width=4cm]{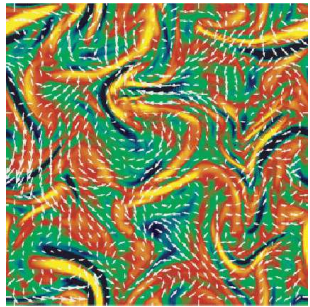}\\
  \caption{Cross section of the current density $J_z$ along the external magnetic field in color tones. Yellow (light) is positive $J_z$, blue (dark) is negative, and the superposed arrows represent the velocity field.  }\label{curturb}
\end{figure}

The magnetic field is directly obtained from the numerical solution of the MHD equations, with  electric field derived from Ohm's law.  It is obvious that the electric field is an intermittent quantity with the high values distributed in a less space filling way. Magnetic and electric fields show a broad range of scales and high degree of complexity.  The energy spectrum of the MHD fields is consistent with a Kolmogorov-5/3 power law. The structure of the velocity  field and the current density along the external magnetic field ($J_z$) can be seen in Fig. \ref{curturb}.
The formation of strong anisotropies in the magnetic field, the fluid velocity and the associated electric field is observed.  The overall picture is that current sheet structures along the DC field are formed as a natural evolution of the MHD fields.

Following thousands of particles particles inside the simulation box, we can learn many of the statistical properties of their evolution, e.g.\ the mean square displacements $\sqrt{<\Delta x^2>}$, $\sqrt{<\Delta v^2>},$  or the velocity distribution etc. can be determined. Electrons and ions are accelerated rapidly 
at the nonlinear small scale structures formed inside the turbulent volume, and non-thermal tails of power-law shape are formed in the velocity distributions. Most particles seem to escape the volume by crossing only a few of the randomly appearing current sheets. A few particles are trapped in these structures and accelerated to very high energies. 
The Fokker-Planck equation is not the appropriate 
tool to capture 
particle motion in the presence of the random appearance of coherent structures 
inside such a 
turbulent environment.

\citet{Vlahos04} performed a Monte Carlo simulation of the extended CTRW
in position and momentum space, in application to flares in the solar
corona, with particular interest in the appearance of the non-thermal
energy distributions of the so-called solar energetic particles.

\section{Summary and Discussion}

Brownian motion is a prototype of normal diffusion, and its analysis has brought
forth a number of tools
that today are very much in use for modeling a wide
variety of phenomena. Normal diffusion occurs in systems which are close to equilibrium,
like the water in Brown's experiment. It has now become evident that phenomena of anomalous
diffusion are very frequent, because many systems of interest are far from equilibrium,
such as turbulent systems,
or because the space accessible to the diffusing particles has a strange, e.g.\ fractal structure.
The tools to model these phenomena, continuous time random walk, stochastic differential
equations, and fractional diffusion equations, are still active research topics.





\begin{thebibliography}{}


\bibitem[Balescu(1995)]{Balescu95}
Balescu, R., Phys.\ Rev. E {\bf 51}, 4807 (1995).

\bibitem[Balescu(2007a)]{Balescu07}
Balescu, R., Chaos Solitons \& Fractals {\bf 34}, 62 (2007).

\bibitem[Balescu(2007b)]{Balescu07b}
Balescu, R., arXiv:0704.2517v1, (2007).

\bibitem[Benisti \& Escande(1998)]{BeEs_98}
Benisti D., and Escande, D.F., Phys. Rev. Lett. \textbf{80}, 4871 (1998).

\bibitem[Blumen, Zumofen \& Klafter(1989)]{Blumen89}
Blumen, A., Zumofen, G., Klafter, J., Phys.\ Rev. A {\bf 40}, 3964 (1989).

\bibitem[Cary, Escande \& Verga(1990)]{CaEsVe_90}
Cary, J.R., Escande D.F. and Verga, A.D. ,  Phys. Rev. Lett. \textbf{65}, 3132 (1990).

\bibitem[Dmitruk et al.(2003)]{Dmitr1}
P. Dmitruk, W.H. Matthaeus, N. Seenu, M.R. Brown: Astrophys. J. Lett., \textbf{597}, L81 (2003)

\bibitem[Dmitruk et al.(2004)]{Dmitr2}
P. Dmitruk, W.H. Matthaeus, N. Seenu: Astrophys. J., \textbf{617}, 667 (2004)

\bibitem[Einstein(1905)]{Einstein05}
Einstein, A., Ann.\ Phys.\ {\bf 17}, 549 (1905)

\bibitem[Escande \& Sattin(2007)]{Escande07}
Escande, D.F \& Sattin, F., Phys.\ Rev.\ Lett. {\bf 99}, 185005 (2007).

\bibitem[Feller(1971)]{Feller71}
Feller, W., An Introduction to Probability Theory and its Applications,
Vol.\ 2, 2nd edition, Wiley (New York), 1971.

\bibitem[Gardiner(2004)]{Gardiner04}
Gardiner, C.W., Handbook of Stochastic Methods, 3rd ed.
(Springer; Berlin \& Heidelberg; 2004)

\bibitem[Goldstein(1980)]{Goldstein} H. Goldstein, Classical Mechanics (2nd edition, Addison-Wesley, Massachusets, 1980).

\bibitem[Helander \& Kjellberg(1994)]{HeKj_94}
Helander P., and Kjellberg, L., Phys. Plasmas \textbf{1}, 210 (1994).

\bibitem[Isliker(2008)]{Isliker08}
Isliker, H., arXiv:0710:5152, and submitted to Phys.\ Rev.\ E (2008)

\bibitem[Hughes(1995)]{Hughes95}
Hughes, B.D., Random Walks and Random Environments, Volume 1,
Random Walks, Oxford (Clarendon Press), 1995.

\bibitem[van Kampen(1981)]{VanKampen}
N.G. van Kampen, Stochastic processes in physics and chemistry (North-Holland, Amsterdam, 1981).

\bibitem[Klafter, Blumen \& Shlesinger(1987)]{Klafter87}
Klafter, J., Blumen, A., Shlesinger, M.F., Phys.\ Rev.\ A {\bf 35}, 3081
(1987).

\bibitem[Kominis(2008)]{PRE}
Kominis, Y.,  \ Phys. \ Rev. E \textbf{77}, 016404 (2008).

\bibitem[Langevin(1908)]{Langevin08}
Langevin, P., C.\ R. Acad.\ Sci.\ (Paris) {\bf 146}, 530 (1908)
[translated to English and commented in Lemons, D.S, Gythiel, A.,
Am.\ J.\ Phys.\ {\bf 65}, 1079 (1997)]

\bibitem[Laval \& Pesme(1983)]{LaPe_83b}
Laval, G.  and Pesme, D.,  Phys. Fluids \textbf{26}, 66 (1983).

\bibitem[Laval \& Pesme(1999)]{LaPe_99}
G. Laval and D. Pesme, Plasma Phys. Control. Fusion \textbf{41}, A239 (1999).

\bibitem[Lenzi, Mendes \& Tsallis(2003)]{Lenzi03}
Lenzi, E.K., Mendes, R.S \& Tsallis, C., Phys.\ Rev.\ E {\bf 67}, 031104 (2003).

\bibitem[Lichtenberg \& Lieberman(1992)]{Lichtenberg92}
Lichtenberg, A. J. and Lieberman, M. A., “Regular and Chaotic Dynamics” (Springer-Verlag, New York, 1992).

\bibitem[Metzler \& Klafter(2000)]{Metzler00}
Metzler, R., Klafter, J., Physics Reports {\bf 339}, 1 (2000).

\bibitem[Metzler \& Klafter(2004)]{Metzler04}
Metzler, R., Klafter, J., Journ.\ of Phys. A {\bf 37}, R161 (2004).

\bibitem[van Milligen, S\'anchez \& Carreras(2004)]{vanMil04}
van Milligen, B.\ Ph., S\'anchez, R., Carreras, B.A., Phys. of Plasmas
{\bf 11}, 2272 (2004)

\bibitem[van Milligen, Carreras \& S\'anchez(2004)]{vanMil04b}
van Milligen, B.\ Ph., Carreras, B.A., S\'anchez, R., Physics of Plasmas
{\bf 11}, 3787 (2004).

\bibitem[van Milligen et al.(2005)]{Milligen05}
van Milligen, B.\ Ph., Bons, P.D., Carreras, B.A., S\'anchez, R.,
Eur.\ J.\ Phys.\ {\bf 26}, 913 (2005).

\bibitem[Molvig \& Hizanidis(1984)]{Molvig84}
Molvig, K., and Hizanidis, K., 'Transport theory: Microscopic reversibility
and symmetry', Phys.\ Fluids {\bf 27}, 2847-2858 (1984).

\bibitem[Montroll \& Weiss(1965)]{Montroll65}
E.W.\ Montroll, G.H.\ Weiss, J.\ Math.\ Phys.\ {\bf 6}, 167 (1965).

\bibitem[Podlubny(1999)]{Podlubny99}
Podlubny, I., Fractional differential equations, New York (Academic press),
1999.

\bibitem[Shlesinger, West \& Klafter(1987)]{Shlesinger87}
Shlesinger, M.F., West, B.,  Klafter, J., Phys. Rev. Lett. {\bf 58}, 1100 (1987).

\bibitem[Solomon, Weeks \& Swinney(1994)]{Solomon94}
Solomon, T.H. Weeks, E.R., Swinney, H.L., Physica D {\bf 76}, 70 (1994).

\bibitem[Vlahos, Isliker \& Lepreti(2004)]{Vlahos04}
Vlahos, L., Isliker, H., Lepreti, F., Astrophys.\ Journ. {\bf 608}, 540 (2004).

\bibitem[Weeks, Urbach \& Swinney(1996)]{Weeks96}
Weeks,E.R., Urbach, J.S., Swinney, H.L., Physica D {\bf 97}, 291 (1996).

\bibitem[Zaslavski \& Filonenko(1968)]{Zaslavski}
G.M. Zaslavski and N.N. Filonenko, Stochastic instability of trapped particles and conditions of applicability of the quasi-linear approximation, Sov. Phys. JETP \textbf{25}, 851 (1968).

\bibitem[Zumofen \& Klafter(1993)]{Zumofen93}
Zumofen, G., Klafter, J., Phys.\ Rev.\ E {\bf 47} 851 (1993).



























\end{thebibliography}
\end{document}